\documentclass[%
 reprint,
nofootinbib,
nobibnotes,
bibnotes,graphicx,graphics,
 amsmath,amssymb,
 aps,stmaryrd,
]{revtex4-1}
\usepackage{natbib} 
\usepackage{xtab,afterpage}
\usepackage{amsfonts}
\usepackage{soul}
\usepackage{tikz}
\newcommand{\comment}[1]{}
\usepackage{tikz}
  \newlength\squareheight
\usepackage{wasysym} 
\usepackage{graphicx}
\usepackage{dcolumn}
\usepackage{bm}

\begin{document}

\preprint{Draft}

\title{Degree correlations in graphs with clique clustering}

\author{Peter Mann}
\email{pm78@st-andrews.ac.uk}
\author{V. Anne Smith}%
\author{John B.O. Mitchell}
\author{Simon Dobson}
\affiliation{School of Computer Science, University of St Andrews, St Andrews, Fife KY16 9SX, United Kingdom }
\affiliation{School of Chemistry, University of St Andrews, St Andrews, Fife KY16 9ST, United Kingdom }
\affiliation{School of Biology, University of St Andrews, St Andrews, Fife KY16 9TH, United Kingdom }

\date{March 29, 2022}

\begin{abstract}
Correlations among the degrees of vertices in random graphs often occur when clustering is present. In this paper we define a joint-degree correlation function for vertices in the giant component of clustered configuration model networks which are comprised of clique subgraphs. We use this model to investigate, in detail, the organisation among nearest-neighbour subgraphs for random graphs as a function of subgraph topology as well as clustering. We find an expression for the average joint degree of a neighbour in the giant component at the critical point for these networks. Finally, we introduce a novel edge-disjoint clique decomposition algorithm and investigate the correlations between the subgraphs of empirical networks.
\end{abstract}

\pacs{Valid PACS appear here}
\maketitle


\section{Introduction}
\label{sec:introduction}

A network is a collection of vertices and edges \cite{newman_2019}. The nature of the local connectivity among the vertices of a graph has a profound influence on the structural characteristics of the entire network. Common structural properties include: the clustering \cite{PhysRevE.72.036133}, which is the tendency for triples of vertices to be organised into triangles; subgraph composition \cite{Arenas_2008}, which considers the organisation of the edges into recognized motifs; nearest-neighbour degree correlation (NNDC) \cite{PhysRevE.98.062314}, which is the tendency for similar degree vertices to connect to one another or not; long-range degree correlations (LRDC) \cite{PhysRevE.97.062308}, which are nonlocal degree correlations beyond the nearest-neighbourhood; the component structure \cite{PhysRevE.76.045101}, the core-periphery structure, path lengths, communities, fractality and various scale phenomena. In turn, the structural characteristics determine the stability and the governing dynamics of processes occurring over the graph as well as its response to random or targeted attack. Understanding the connective microstructure of complex systems is therefore of crucial importance to a wide range of disciplines including biology, social science and physics as well as to a broad range of applications including network formation, modelling the properties of empirical networks and the observed response to processes such as epidemic spreading, synchronization, percolation or information propagation over networks. It is well known \cite{PhysRevE.77.036124,PhysRevE.97.042318,PhysRevE.99.042308} that the structural characteristics of the giant component (GCC) of a random uncorrelated graph can be vastly different from the properties of the whole network. In particular, the GCC exhibits a negative NNDC unless the network is singly connected.   

The configuration model is a method that allows the construction of uncorrelated random graphs with a prescribed distribution of degrees. Recent work has drawn attention to the generalised configuration model (GCM) which allows the construction of networks that are composed of independent subgraphs. The central object of the GCM is a joint degree distribution that describes the number of roles that a vertex plays within each subgraph on average \cite{karrer_newman_2010,PhysRevE.103.012313,PhysRevE.103.012309}. The generating function formulation is an analytical technique that can be used to describe the expectation values for the properties of the ensemble of graphs that can be constructed using the GCM from a given joint degree sequence. 

The GCM incorporates networks with higher-order clustering, typical of the mixing patterns in many human contact networks, as well as multilayer, modular and multiplex systems. In such empirical networks, clustering that follows a heavy tail degree distribution leads to highly clustered networks whereby the vertices can be members of several triangles among the nearest-neighbour contacts. In such cases, it is common that the triangles share one or more edges and thus,  higher-order subgraphs, such as cliques, are more accurate representations of the local environment of the vertices. Organisation among cliques of different sizes plays a significant and non-trivial role in spreading processes, particularly of epidemics, over the network. Since many diseases spread through vertex-vertex interactions, effective control of an epidemic must take advantage of the understanding of the local environment of high-degree vertices in tight-knit cliques. 

Clustering in complex networks has been studied previously using generating functions \cite{PhysRevE.68.026121,miller_2009_spread,PhysRevE.80.020901,karrer_newman_2010,gleeson_2009,PhysRevE.83.056107,PhysRevE.81.066114,allard_hebert-dufresne_young_dube_2015,PhysRevE.103.012313,PhysRevE.103.012309,Peron_2018,HASEGAWA2021125970,PhysRevE.101.062310,PhysRevE.104.024304,Stegehuis_2021}. Newman \cite{PhysRevLett.103.058701} found that the presence of clustering in Poisson networks led to a reduction in the critical mean degree required for the formation of a GCC as well as its size. Miller \cite{PhysRevE.80.020901} showed that this effect is due to the assortative correlations within the Poisson model and that for networks with the same degree correlations, clustering increases the critical point. Hasegawa and Mizutaka \cite{PhysRevE.101.062310} considered the NNDC among the GCC of clustered networks comprised of ordinary edges and triangles. It was found that the GCC can be assortative or dissasortative depending on the details of the clustering; however, dissasortative correlations reappeared upon a characteristic renormalisation of the triangles into single \textit{supervertices}. Thus, the GCC of random uncorrelated networks displays dissasortative NNDC by nature.

In this paper, we address how two vertices of given joint degrees are expected to connect to one another. More formally, we study NNDC in the GCC of random clustered graphs that have been constructed according to the GCM prescription to include higher-order subgraphs. We examine the tendency for organisation among the subgraphs and investigate whether vertices with high subgraph degree connect preferentially to other high subgraph degree vertices or not. We then examine the properties of empirical networks by introducing a novel clique cover and compare our cover to other recent advances in the literature \cite{2021arXiv210103618B}.

\section{Background}
\label{sec:background}
In this section, we review the generating function formulation for higher-order subgraphs \cite{PhysRevLett.103.058701,PhysRevE.80.020901,karrer_newman_2010,PhysRevE.103.012313,PhysRevE.103.012309} and the method of construction of GCM networks. We reserve bold characters for vector quantities.

The degree distribution $p_k$ is the probability that a randomly chosen vertex in the network has degree $k$. A common assumption is that the edges are locally tree-like; short range cycles and connections among the nearest neighbours are prohibited. The tree-like assumption has proven very successful at describing many network properties \cite{PhysRevE.83.036112}; however, the properties of random clustered networks require a generalisation to the degree of a vertex, beyond simple tree edges, to incorporate the effects of triangles and other higher-order motifs. The resulting model was developed independently by Newman \cite{PhysRevLett.103.058701} and Miller \cite{PhysRevE.80.020901} for networks with triangles and later extended to all network motifs by Karrer and Newman \cite{karrer_newman_2010}. The models assume that overall degree of a vertex can be partitioned into sub-degrees that correspond to the involvement of a vertex in pre-defined subgraphs. For instance, the generalised degree, $\bm {k_\tau}=(k_\bot,k_{\Delta},k_{\square},\dots)$, of a vertex that has six tree-like edges and is also a member of one triangle, two squares and three pentagons would be $\bm {k_\tau}=(6,1,2,3)$. The probability that a randomly chosen vertex has a particular generalised degree is given by a joint degree distribution $p_{\bm {k_\tau}}$. The ordinary degree distribution is recovered from 
\begin{equation}
    p_k=\sum^\infty_{k_\bot=0}\cdots\sum^\infty_{k_\gamma=0}
    p_{k_\bot,\dots,k_\gamma}\delta_{k,\sum \lambda_\tau k_{\tau\in\bm{\tau}}}
\end{equation}
\noindent where $\bm{\tau}$ is a vector of subgraph topologies $\{\bot, \triangle, \square, \pentagon, \cdots ,\gamma\}$, up to some terminating motif topology represented by $\gamma$, $k_\tau$ is the degree of shape $\tau\in \bm \tau$, $\lambda_\tau$ is the number of edges a vertex has in shape $\tau$, $p_{\bm{k_\tau}}=p_{k_\bot,\dots,k_\gamma}$ is the $\dim(\bm\tau)$ joint probability distribution of degrees and $\delta_{i,j}$ is the Kronecker delta. For instance, a vertex that is part of a two tree-like edges, a triangle and a square will have the following joint degree sequence $(k_\bot,k_\triangle,k_\square)=(2,1,1)$, while its overall degree is $k=6$. A network is described by its joint probability distribution of each vertex playing a certain role in a given subgraph a particular number of times \cite{karrer_newman_2010} for all permissible combinations of joint degrees. The joint degree distribution can be generated using 
\begin{align}
     G_0(\bm{z})=& \sum^\infty_{k_\bot=0}\cdots \sum_{k_\gamma=0}^\infty p_{k_\bot,\dots,k_\gamma}z_\bot^{k_\bot}\cdots z_\gamma^{k_{\gamma}}\label{eq:G0}
\end{align}
\noindent where $\bm z = \{ z_\bot, z_\triangle, z_\square, \dots ,z_\gamma\}$. In the ordinary generating function model, the excess degree distribution $q_k$ defines the probability that a randomly chosen edge leads to a vertex of degree $k+1$. In the generalised model we must define an excess degree distribution for each topology in $\bm \tau$; since, traversing an edge of a particular topology does not, in general, lead to vertices with equivalent joint degrees. The joint excess degree distribution for an edge of topology $\tau$ is 
\begin{equation}
    q_\tau(\bm {k_\tau})={(k_\tau+1)p_{k_{\bm\tau\backslash\{\tau\}}, k_{\tau}+1}}/{\langle k_\tau\rangle}
\end{equation}
where the notation $\bm {s}\backslash\{s\}$ excludes element $s$ from set $\bm {s}$. Each joint excess degree distribution is generated as
\begin{align}
    G_{1,\tau}(\bm z) =&\sum^\infty_{k_\bot}\cdots \sum^\infty_{k_\gamma}q_{\bm{k_\tau}}\
    z_\tau^{k_\tau-1}\prod_{\nu\neq \tau} z_\nu^{k_{\nu}}
\end{align}
\noindent and is also seen to be the partial derivative of Eq. \ref{eq:G0} with respect to $z_\tau$ divided by the expected number of $\tau$-motifs
\begin{equation}
    G_{1,\tau}(\bm z) = \frac 1{\langle k_\tau\rangle}  \frac{\partial G_0}{\partial z_\tau}\label{eq:Jacobian}
\end{equation}
which can also be written as 
\begin{equation}
    G_{1,\tau}(\bm z) = \frac{G_0^{'\tau}(\bm z)}{G_0^{'\tau}(\bm 1)}
\end{equation}
where $G_0^{'\tau}$ is the first derivative of $G_0(\bm z)$ with respect to $z_\tau$ and $\langle k_\tau\rangle=G_0^{'\tau}(\bm 1)$ is the average $\tau$-degree for a vertex in the network. 

The global clustering coefficient $C$ of a network with $V$ vertices is defined as
\begin{equation}
    C=\frac{3\mathcal N_\Delta }{\mathcal N_3}
\end{equation}
where $\mathcal N_\Delta$ is the number of triangles in the network and $\mathcal N_3$ is the number of connected triples. The number of triangles involving vertices with a given joint degree $\bm k_{\bm \tau}$ is 
\begin{equation}
    \mathcal N_{\Delta,\bm k_{\bm \tau}} = V
    p_{{k_\bot,\dots,k_\gamma}}(k_\Delta + \dots + \mu_\gamma k_\gamma)\label{eq:clu}
\end{equation}
where $\mu_\tau$ is the number of triangles that a vertex belongs to as a member of a $\tau$-motif. For instance, $\mu_\Delta=1$ while a vertex in 4-clique has belongs to 3 triangles. The total number of triangles in the network is found by summing over the joint degree
\begin{equation}
    \mathcal N_{\Delta}= \sum^\infty_{k_\bot=0}\cdots\sum^\infty_{k_\gamma=0}\mathcal N_{\Delta,\bm k_{\bm \tau}}
\end{equation}
The number of connected triples is given by \cite{PhysRevLett.103.058701}
\begin{equation}
    \mathcal N_3 = V\sum_k \binom{k}{2}p_k
\end{equation}
We can use the generating function formulation to determine the probability that a vertex selected at random belongs to the GCC. Let $u_\tau$ be the probability that a vertex reached by the traversal of an edge of topology $\tau$ does not lead to the GCC. Similarly, the probability that the entire subgraph does not connect the vertex to the GCC is $u_\tau^{m_\tau}$ where $m_\tau$ is the number of edges a vertex has in each independent subgraph of topology $\tau$. For instance, a vertex has 3 edges in a given 4-clique. The probability that the neighbour fails to attach to the GCC is given by a self-consistent expression $u_\tau=G_{1,\tau}(\bm {u_\tau^{m_\tau}})$ where $\bm {u_\tau^{m_\tau}}=\{u_\bot,u_\Delta^2, \dots,u_\gamma^{m_{\gamma}}\}$. The size of the largest percolating cluster $S$ can then be calculated as
\begin{equation}
    S = 1 - G_0(\bm {u_\tau^{m_\tau}})\label{EQ:main}
\end{equation}
Introducing $H(\bm x)$ as the generating function for the GCC as 
\begin{equation}
    H(\bm x) = \frac{G_0(\bm x) - G_0(\bm x \cdot\bm {u_\tau^{m_\tau}})}{1 - G_0(\bm { u_\tau^{m_\tau}})}
\end{equation}
where $\bm v\cdot \bm w$ is the scalar product $v_iw_i$. The overall degree distribution of the GCC is given by 
\begin{equation}
    p^{\text{GCC}}_k = \frac{1}{k!}\frac{\partial ^k}{\partial x^k}H(\bm {x^{m_\tau}})\bigg|_{\bm x=\bm 1}
\end{equation}
where $\bm x = (x,x^2,\dots,x^{m_\gamma})$. The networks that we use in this paper are constructed according to the GCM which we now detail \cite{fosdick_larremore_nishimura_ugander_2018, 10.1093/comnet/cnw011,wang_lizardo_hachen_2014,heath_parikh_2011}. For each vertex in a collection of vertices, a joint degree is chosen from a distribution of joint degrees to create a joint degree sequence. Not all joint degree sequences are valid or \textit{graphic} \cite{PhysRevE.104.024304}. There is a constraint on the permissible sequence of joint degrees generated such that the sum of the number of motifs of each kind is divisible by the number of vertices in each basis motif. For instance, the number of triangles in the joint degree sequence must be divisible by 3 and so on. This ensures that when the vertices are chosen at random and connected, there are precisely the correct number of edges to construct each motif. This constraint does not impact the number of each motif in the network; however.

Once the vertices have been assigned their stub degrees, they are connected at random to form the appropriate subgraphs according to their joint degree sequence through a stub-matching process. The probability of accidental formation of short range loops or motifs that share edges (non-edge disjoint motifs) becomes vanishingly small in the limit that the networks are large. Upon renormalising each motif to its characteristic scale based on neighbouring vertex count, we recover the treelike property of the original configuration model.

\section{Theoretical}
\label{sec:NNHO}

Consider an arbitrary set of edge topologies including ordinary edges, triangles, squares, 4-cliques, pentagons and so on, denoted by $\bm {\vec \tau} = \{\bot,\Delta,\square, \dots, \gamma\}$, where $\gamma$ is the topology of the final element. In the following, we reserve $\tau$ and $\nu$ as indices over elements of $\bm \tau$. We define the number of subgraphs that a vertex plays a role in for each topology $\tau\in \bm \tau$ by vector ${\bm{k}_{\bm{\tau},l}} = \{k_\bot,k_\Delta,\dots, k_{\gamma}\}$ with $l=0,1$ representing the focal vertex and nearest-neighbour joint sequences, respectively. We reserve $k_{\nu,l}\in {\bm{k}_{\bm{\tau},l}}$ as an index for the number of subgraphs of topology $\nu$ around a given vertex in layer $l$; we drop the $l$ label where obvious. The joint probability distribution for choosing this vertex at random is then denoted as $p_{\bm{k}_{\bm{\tau},l}}$. The number of edges that a given vertex has within each motif is defined by $m_\tau$; for instance a vertex contributes two edges to each triangle it connects to and hence $m_\Delta=2$.

We define $n_{\tau,\nu,k_\nu}$ to be the number of vertices with $k_\nu$ subgraphs of topology $\nu$ that we reach by following an edge of topology $\tau$ from the focal vertex to a nearest neighbour. There are $\dim{(\bm \tau}^2)$ of these expressions. Let a particular configuration of type $\nu$ following $\tau$ edges be $n_{\tau,\nu}$ such that 
\begin{equation}
    n_{\tau,\nu} = \{ n_{\tau,\nu,1},n_{\tau,\nu,2},\dots\}
\end{equation}
\begin{figure}
\begin{center}
\includegraphics[width=0.475\textwidth]{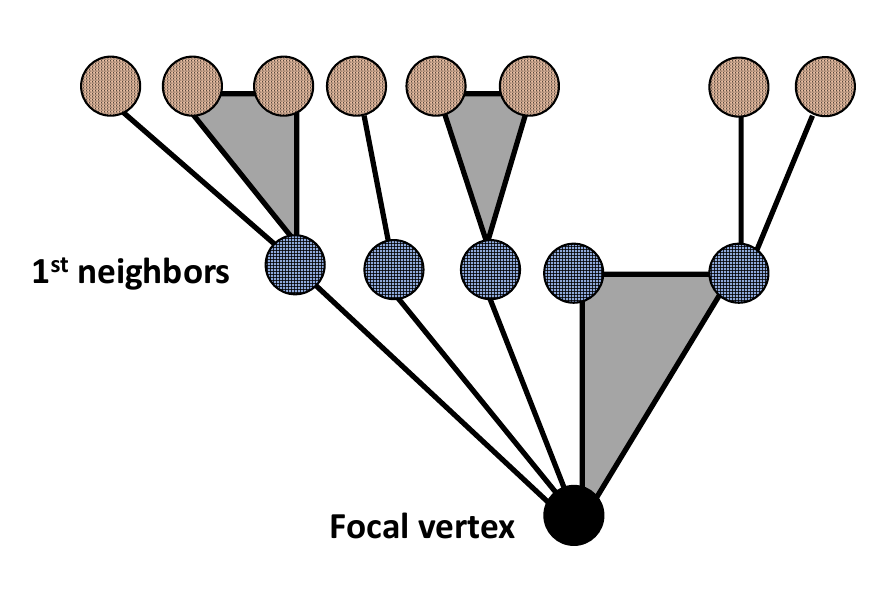}
\caption[countingshapes]{ 
A focal vertex in a 2- and 3-clique random graph with $n_{\bot,\bot,1}=n_{\bot,\Delta,0}=n_{\Delta,\bot,0}=n_{\Delta,\bot,2}=1$ and $n_{\bot,\bot,2}=n_{\bot,\Delta,1}=n_{\Delta,\Delta,1}=2$.
} \label{fig:countingStuff}
\end{center}
\end{figure}
For instance, for a focal vertex that belongs to a GCM graph comprising of vertices with both 2- and 3-cliques such that $\bm \tau=\{\bot,\Delta\}$, the configuration of 3-cliques obtained by following 2-cliques to a neighbour is 
\begin{equation}
    n_{\bot,\Delta} = \{n_{\bot,\Delta,1},n_{\bot,\Delta,2},\dots,n_{\bot,\Delta,k_{\Delta, \text{max}}}\}
\end{equation}
where $k_{\Delta, \text{max}}$ is the maximum number of triangles a single vertex belongs to, see Fig \ref{fig:countingStuff}.

Then, we define the set of all configurations of the neighbours following $\tau$ edges to be $n_\tau = \{n_{\tau,\bot}, n_{\tau,\Delta},\dots\}$. For instance, returning to the mixed 2- and 3-clique example, we can also count the number of 2-cliques the neighbour has instead of enumerating the 3-cliques. Therefore, for this example we have 
\begin{equation}
    n_\bot = \{n_{\bot,\bot}, n_{\bot,\Delta}\}
\end{equation}
Finally, the set of all configurations of neighbour motif membership is denoted by $n = \{n_\bot,n_\Delta,\dots\}$, which accounts for each edge-type we could have followed to reach the neighbour vertices. 

The number of vertices reached by following all of the $\tau$ edges is 
\begin{equation}
    N_\tau= \sum_{k_\tau=1}n_{\tau,\tau,k_\tau}=\sum_{\nu=0}n_{\tau,\nu,k_\nu}\qquad \tau\neq\nu
\end{equation}
For instance, for the focal vertex in Fig \ref{fig:countingStuff}, we have \begin{equation}
    \sum_{k_\bot=1}n_{\bot,\bot,k_\bot}=\sum_{k_\Delta=0}n_{\bot,\Delta,k_\Delta}=3
\end{equation} 
and 
\begin{equation}
    \sum_{k_\Delta=1}n_{\Delta,\Delta,k_\Delta}=\sum_{k_\bot=0}n_{\Delta,\bot,k_\bot}=2
\end{equation} 
The total number of vertices 1-layer out from the focal vertex is the sum of all vertices reached by traversing each edge topology 
\begin{equation}
    N = \sum_{\tau\in \bm \tau}N_{\tau}
\end{equation}
and hence, for the focal vertex in Fig \ref{fig:countingStuff}, the total number of direct neighbours is given by $N=5$.

Let $P(n\mid N)$ be the probability that the nearest-neighbour configuration is given by set $n$ and that the total number of vertices in the first layer is $N$. This is given by 
\begin{widetext}

\begin{equation}
    P(n\mid N) = \prod_{\tau}\bigg(\prod_{\nu\neq \tau}\prod\limits_{k_\nu=0}\frac{N_\tau}{n_{\tau,\nu,k_\nu}!} q_{\tau,\nu,k_\nu}^{n_{\tau,\nu,k_\nu}}\bigg)\prod\limits_{k_\tau=1}\frac{N_\tau}{n_{\tau,\tau,k_\tau}!}q_{\tau,\tau,k_\tau}^{n_{\tau,\tau,k_\tau}}
\end{equation}
where $q_{\tau,\nu,k}$ is the probability of traversing an edge of topology $\tau$ to a vertex with $k_\nu$ independent subgraphs of topology $\nu$. We also have the understanding that each term of the product over $\nu\neq \tau$ has its own index $k_\nu$ starting from zero; we have pulled out $\tau$ from this expression since, by definition, there must be at least one $\tau$-edge present to follow it to a nearest neighbour vertex and so the index starts at 1. The probability $P(\text{GCC}\mid n)$ that the component is the GCC for a particular configuration $n$ is given by 
\begin{equation}
 P(\text{GCC}\mid n, N) = 1- \prod_{\tau}\bigg(\prod_{\nu\neq \tau}\prod_{k_\nu=0}\big(u_\nu^{m_\nu k_\nu}\big)^{n_{\tau,\nu,k_\nu}}\bigg)  \prod_{k_\tau=1}(u_\tau^{m_\tau(k_\tau-1)})^{n_{\tau,\tau,k_\tau}}
\end{equation}
where we have introduced $u_\tau$ as the probability that a vertex at the end of a randomly chosen edge of topology $\tau$ fails to connect to the GCC. The probability that the configuration is $n$, that the component is the GCC given that there are $N$ nearest-neighbours is found from Bayes' theorem as
\begin{equation}
    P(n,\text{GCC}\mid N) = P(\text{GCC}\mid n,N) P(n\mid N)
\end{equation}
Let $P(N\mid \bm k_{\bm \tau,0})$  be the probability of there being $N$ vertices in the 1st layer given that the joint degree of the focal vertex is $\bm k_{\bm \tau,0}$ and that the component is the GCC. We can use this to find the probability $P(n,\text{GCC}\mid \bm k_{\bm \tau,0})$ that the nearest-neighbour configuration is $n$ given the joint degree of a vertex in the GCC is $\bm k_{\bm \tau,0}$ as 
\begin{equation}
    P(n,\text{GCC}\mid \bm k_{\bm \tau,0}) = \sum_{N} P(N\mid \bm k_{\bm \tau,0})P(n,\text{GCC}\mid N)
\end{equation}
where the summation is over all combinations of $N_\tau$ such that 
\begin{equation}
    \sum_N =  \sum_{N_\bot}\sum_{N_\Delta}\cdots
\end{equation}
We find
\begin{align}
     P(n,\text{GCC}\mid \bm k_{\bm \tau,0}) =\ &  \sum_{N} P(N\mid \bm k_{\bm \tau,0}) \prod_{\tau}\left(\prod_{\nu\neq\tau}\prod\limits_{k_\nu=0}\frac{N_\tau}{n_{\tau,\nu,k_\nu}!} q_{\tau,\nu,k_\nu}^{n_{\tau,\nu,k_\nu}}\right)\prod\limits_{k_\tau=1}\frac{N_\tau}{n_{\tau,\tau,k_\tau}!}q_{\tau,\tau,k_\tau}^{n_{\tau,\tau,k_\tau}}\nonumber\\
     &\times \left[ 1- \prod_{\eta}\left(\prod_{\varphi\neq \eta}\prod_{k_\nu=0}\left(u_\varphi ^{m_\varphi k_\nu}\right)^{n_{\eta,\varphi,k_\nu}}
     \right)  \prod_{k_\tau=1}
     \left(u_\eta^{m_\eta(k_\tau-1)}\right)^{n_{\eta,\eta,k_\tau}}
     \right]\qquad \tau,\nu,\eta,\varphi\in \bm \tau\label{eq:sub1}
\end{align}
We now generate this probability by summing over all permissible configurations of the nearest-neighbour joint degrees to obtain
\begin{equation}
    \Tilde F_{\text{GCC}}(\bm X\mid \bm k_{\bm \tau,0})=\sum_n P(n,\text{GCC}\mid\bm k_{\bm \tau,0})\prod_\tau\left(\prod_{\nu\neq \tau} \prod_{k_\nu=0}X_{\tau,\nu,k_\nu}^{n_{\tau,\nu,k_\nu}}\right)\prod_{k_\tau=1}X_{\tau,\tau,k_\tau}^{n_{\tau,\tau,k_\tau}}
\end{equation}
where 
\begin{equation}
    \sum_n = \sum_{n_{\bot,\bot}}\sum_{n_{\bot,\Delta}}\cdots \sum_{n_{\Delta,\bot}}\sum_{n_{\Delta,\Delta}}\cdots
\end{equation}
We simplify the expression by substituting Eq \ref{eq:sub1}, swapping the order of the summations and collecting terms in like powers to obtain
\begin{align}
     \Tilde F_{\text{GCC}}(\bm X\mid \bm k_{\bm \tau,0})=\ &\sum_n \sum_{N} P(N\mid \bm k_{\bm \tau,0}) \prod_{\tau}\left(\prod_{\nu\neq\tau}\prod\limits_{k_\nu=0}\frac{N_\tau}{n_{\tau,\nu,k_\nu}!} q_{\tau,\nu,k_\nu}^{n_{\tau,\nu,k_\nu}}\right)\prod\limits_{k_\tau=1}\frac{N_\tau}{n_{\tau,\tau,k_\tau}!}q_{\tau,\tau,k_\tau}^{n_{\tau,\tau,k_\tau}}\nonumber\\
     &\times \left[ 1- \prod_{\eta}\left(\prod_{\varphi\neq \eta}\prod_{k_\nu=0}\left(u_\varphi ^{m_\varphi k_\nu}\right)^{n_{\eta,\varphi,k_\nu}}
     \right)  \prod_{k_\tau=1}\left(u_\eta^{m_\eta(k_\tau-1)}\right)^{n_{\eta,\eta,k_\tau}}\right]\prod_\tau\left(\prod_{\nu\neq \tau} \prod_{k_\nu=0}X_{\tau,\nu,k_\nu}^{n_{\tau,\nu,k_\nu}}\right)\prod_{k_\tau=1}X_{\tau,\tau,k_\tau}^{n_{\tau,\tau,k_\tau}}
\end{align}
to find
\begin{align}
     \Tilde F_{\text{GCC}}(\bm X\mid \bm k_{\bm \tau,0})=\ &\sum_n \sum_{N} P(N\mid \bm k_{\bm \tau,0}) \prod_{\tau}\left(\prod_{\nu\neq\tau}\prod\limits_{k_\nu=0}\frac{N_\tau}{n_{\tau,\nu,k_\nu}!} (q_{\tau,\nu,k_\nu}X_{\tau,\nu,k_\nu})^{n_{\tau,\nu,k_\nu}}\right)\prod\limits_{k_\tau=1}\frac{N_\tau}{n_{\tau,\tau,k_\tau}!}(q_{\tau,\tau,k_\tau}X_{\tau,\tau,k_\tau})^{n_{\tau,\tau,k_\tau}}\nonumber\\
     &\times \left[ 1- \prod_{\eta}\left(\prod_{\varphi\neq \eta}\prod_{k_\nu=0}
     \left(u_\varphi ^{m_\varphi k_\nu}\right)^{n_{\eta,\varphi,k_\nu}}
     \right)  \prod_{k_\tau=1}\left(u_\eta^{m_\eta(k_\tau-1)}\right)^{n_{\eta,\eta,k_\tau}}\right]
\end{align}
The multinomial theorem can now be applied to each of the terms in the product to obtain
\begin{align}
     \Tilde F_{\text{GCC}}(\bm X\mid \bm k_{\bm \tau,0})=\ &\sum_{N} P(N\mid \bm k_{\bm \tau,0}) \prod_{\tau}\left[\left( \prod_{\nu\neq\tau}\sum_{k_\nu=0}q_{\tau,\nu,k_\nu}X_{\tau,\nu,k_\nu}\right)\sum_{k_\tau=1}q_{\tau,\tau,k_\tau}X_{\tau,\tau,k_\tau}\right]^{N_\tau}\nonumber\\
     &- \sum_{N} P(N\mid \bm k_{\bm \tau,0}) \prod_{\tau}\left[\left( \prod_{\nu\neq\tau}\sum_{k_\nu=0}q_{\tau,\nu,k_\nu}u_\nu ^{m_\nu k_\nu}X_{\tau,\nu,k_\nu}\right)\sum_{k_\tau=1}q_{\tau,\tau,k_\tau}u_\tau^{m_\tau(k_\tau-1)}X_{\tau,\tau,k_\tau}\right]^{N_\tau}
\end{align}
The probability that an edge of topology $\tau$ can be followed to reach a vertex with $k_\nu$ subgraphs of topology $\nu$ is given by $q_{\tau,\nu,k_\nu}$. The probability that an edge of topology $\tau$ can be traversed to reach a vertex with $k_\nu$ motifs of topology $\nu$ for all $\nu\in \bm \tau$ is the joint excess degree distribution, $q_{\tau,\bm k_{\bm \tau,l}}$. This can be constructed from the separable distributions such that 
\begin{equation}
    q_{\tau,\bm k_{\bm \tau,l}} = \prod _\nu q_{\tau,\nu,k_{\nu,l}}
\end{equation}
With this we can write
\begin{align}
     \Tilde F_{\text{GCC}}(\bm X\mid \bm k_{\bm \tau,0})=\ &\sum_{N} P(N\mid \bm k_{\bm \tau,0}) \prod_{\tau}\left( \prod_{\nu\neq\tau}\sum_{k_\tau=1}\sum_{k_\nu=0}q_{\tau,\bm k_{\bm \tau},1}X_{\tau,\nu,k_\nu}X_{\tau,\tau,k_\tau} \right)^{N_\tau}\nonumber\\
     &- \sum_{N} P(N\mid \bm k_{\bm \tau,0}) \prod_{\tau}\left( \prod_{\nu\neq\tau}\sum_{k_\tau=1}\sum_{k_\nu=0}q_{\tau,\bm k_{\bm \tau},1}u_\nu ^{m_\nu k_\nu}u_\tau^{m_\tau(k_\tau-1)}X_{\tau,\nu,k_\nu}X_{\tau,\tau,k_\tau}\right)^{N_\tau}
\end{align}
The probability that there are $N$ nearest-neighbour vertices given the joint degree of the focal vertex is $\bm k_{\bm \tau,0}$ is simply a particular term from the $G_0(\bm Z)$ generating function. Inserting this definition into our expression we arrive at the generating function that describes the distribution of nearest-neighbours given a particular joint degree of the focal vertex as 
\begin{align}
    \hat F_{\text{GCC}}(\bm X\mid \bm k_{\bm \tau,0})=&\ p_{\bm k_{\bm \tau,0}} \prod_\tau\left( \prod_{\nu\neq \tau}\sum_{k_\tau=1}\sum_{k_\nu=0}q_{\tau,\bm k_{\bm \tau,1}} X _{\tau,\nu,k_\nu}X_{\tau,\tau,k_\tau}\right)^{m_\tau k_{\tau,0}}\nonumber\\
    &\ - p_{\bm k_{\bm \tau,0}}\prod_{\tau}\left( \prod_{\nu\neq\tau}\sum_{k_\tau=1}\sum_{k_\nu=0}q_{\tau,\bm k_{\bm \tau},1}u_\nu ^{m_\nu k_\nu}u_\tau^{m_\tau(k_\tau-1)}X_{\tau,\nu,k_\nu}X_{\tau,\tau,k_\tau}\right)^{m_\tau k_{\tau,0}}\label{eq:FHAT}
\end{align}
The expectation number of nearest-neighbours with a given joint degree is found from the expectation value of $\hat F_{\text{GCC}}(\bm X=Z\mid \bm k_{\bm \tau,0})$. We then find 
\begin{equation}
   {\hat{F}_{\text{GCC}}'} = \sum_{\tau\in \bm \tau}m_{\tau}p_{\bm{k}_{\bm{\tau},0}}{{k}_{{\tau},0}}q_{\tau,{\bm{k}_{\bm{\tau},1}}}\bigg(1-u_\tau^{m_\tau(k_{\tau,0}+k_{\tau,1}-1)-1}\prod_{\nu\in \bm \tau\backslash \tau}u_\nu^{m_\nu(k_{\nu,0} + k_{\nu,1})}\bigg)
\end{equation}
where the derivative is evaluated at $Z_{\bm {k_{\tau,1}}}=1$ (see Appendix \ref{sec:appendixA} for a complete derivation using the tree-triangle model). The bracket is one minus the probability that the none of the edges to the second layer lead to the GCC; whilst the prefactor describes the probability of following $k_{\tau,0}$ $\tau$-motifs, each of which has $m_\tau$ edges to follow to reach a vertex whose joint degree is given by $q_{\tau,\bm {k_{\tau,1}}}$. The exponent of $u_\tau$ is the number of neighbouring vertices that can be reached by following edges belonging to $\tau$-subgraphs incident to two vertices at the end of an edge in a $\tau$ motif. This is the total number of $\tau$ edges minus the $m_\tau$ that belong to the focal edge's motif minus the focal edge itself.
\begin{equation}
    m_\tau(k_{\tau,0}+k_{\tau,1}-2)+m_\tau-1 = m_\tau(k_{\tau,0}+k_{\tau,1}-1)-1
\end{equation}
In a similar way, we can find the generating function $F_{\text{GCC}}(\bm X)$ for the probability distribution that a randomly chosen vertex has a nearest neighbour configuration given by $n$ and belongs to the GCC as 
\begin{align}
    F_{\text{GCC}}(\bm X) =& \sum_{\bm k_{\bm \tau,0}} {\hat{F}_{\text{GCC}}}(\bm X\mid \bm k_{\bm \tau,0})\\
    =&\sum_{\bm k_{\bm \tau,0}} p_{\bm k_{\bm \tau,0}} \prod_\tau\left( \prod_{\nu\neq \tau}\sum_{k_\tau=1}\sum_{k_\nu=0}q_{\tau,\bm k_{\bm \tau,1}} X _{\tau,\nu,k_\nu}X_{\tau,\tau,k_\tau}\right)^{m_\tau k_{\tau,0}}\nonumber\\
    &\ - \sum_{\bm k_{\bm \tau,0}} p_{\bm k_{\bm \tau,0}}\prod_{\tau}\left( \prod_{\nu\neq\tau}\sum_{k_\tau=1}\sum_{k_\nu=0}q_{\tau,\bm k_{\bm \tau},1}u_\nu ^{m_\nu k_\nu}u_\tau^{m_\tau(k_\tau-1)}X_{\tau,\nu,k_\nu}X_{\tau,\tau,k_\tau}\right)^{m_\tau k_{\tau,0}}\label{eq:F}
\end{align}
\begin{figure}
\begin{center}
\includegraphics[width=0.999\textwidth]{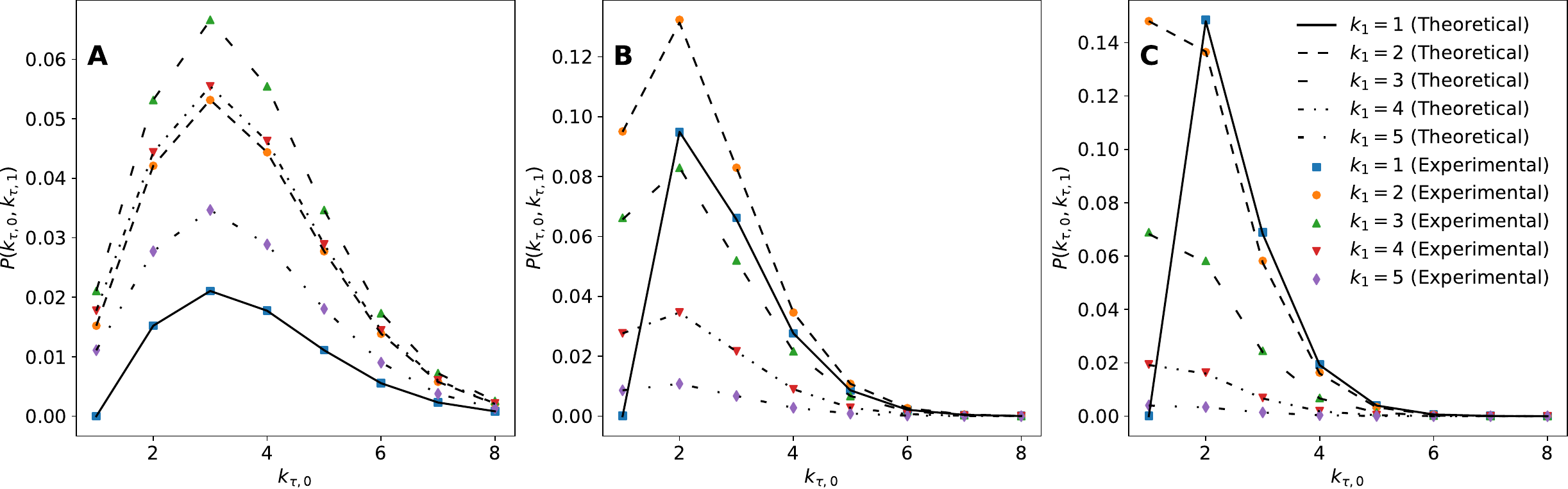}
\caption[cliquePKKsingle]{ 
The probability $P(k_{\tau,0},k_{\tau,1})$ for Poisson random graphs comprising of a single motif topology, 2-cliques (A), 3-cliques (B) and 4-cliques (C), respectively, as a function of $k_{\tau,0}$ for several $k_{\tau,1}$. The overall mean degree is fixed at $\langle k\rangle=2.5$ for networks with $N=60000$ vertices. Scatter points are the average of 100 repetitions of Monte Carlo simulation while the lines are the theoretical predictions from Eq \ref{eq:singletopologymain}. The legend is the same for each plot. 
} \label{fig:cliquePKKsingle}
\end{center}
\end{figure}

\noindent which is simply $G_0(\bm Z)$. The expectation number for the of nearest-neighbours from a random focal vertex in the GCC is given by 
\begin{equation}
   F_{\text{GCC}}' = \sum_{\tau\in \bm \tau}m_\tau \langle k_\tau\rangle \left(1-u_\tau^{m_\tau \omega_\tau}\right)
\end{equation}
where $\omega_\tau$ represents the number of vertices in the motif. We can use the quotient of these expectation values to define a symmetric joint-probability distribution $P_{\text{GCC}}(\bm k_{\bm \tau,0},\bm k_{\bm \tau,1})= \hat{F}_{\text{GCC}}'/F_{\text{GCC}}'$ that two nearest-neighbours in the GCC have joint degrees $\bm k_{\bm \tau,0}$ and $\bm k_{\bm \tau,1}$ as
\begin{equation}
    P_{\text{GCC}}(\bm k_{\bm \tau,0},\bm k_{\bm \tau,1}) =\sum_{\tau\in \bm \tau}m_{\tau}p_{\bm{k}_{\bm{\tau},0}}{{k}_{{\tau},0}}q_{\tau,{\bm{k}_{\bm{\tau},1}}}\bigg(1-u_\tau^{m_\tau(k_{\tau,0}+k_{\tau,1}-1)-1}\prod_{\nu\in \bm \tau\backslash \tau}u_\nu^{m_\nu(k_{\nu,0} + k_{\nu,1})}\bigg)/\sum_{\tau\in \bm \tau}m_\tau \langle k_\tau\rangle \left(1-u_\tau^{m_\tau \omega_\tau}\right)\label{eq:central}
\end{equation}
where $P_{\text{GCC}}(\bm k_{\bm \tau,0},\bm k_{\bm \tau,1})=P_{\text{GCC}}(k_{\bot,0},\dots,k_{\gamma,0},k_{\bot,1},\dots,k_{\gamma,1})$. This equation is a central result and can be used to compute many interesting properties of the correlation structure within configuration model networks. At any time, we can compress the information contained within  $P_{\text{GCC}}(\bm k_{\bm \tau,0},\bm k_{\bm \tau,1})$ to find $P_{\text{GCC}}(k_0,k_1)$ which is the probability that a focal vertex with overall degree $k_0$ attaches to a neighbour whose overall degree is $k_1$. 
\begin{equation}
    P^{\text{overall}}_{\text{GCC}}(k_0,k_1)=\sum_\tau\sum_{k_\tau}  P_{\text{GCC}}(\bm k_{\bm \tau,0},\bm k_{\bm \tau,1}) \delta _{k_0,k_0^{\text{overall}}}\delta_{k_1,k_1^{\text{overall}}}\label{eq:lumpeddegrees}
\end{equation}
where $k_0^{\text{overall}}=\sum_{\tau}\sum_{k_{\tau,0}} m_\tau k_{\tau,0}$ and  $k_1^{\text{overall}}=\sum_{\tau}\sum_{k_{\tau,1}} m_\tau k_{\tau,1}$ are the overall degrees of the focal and neighbour vertices. However, this degree lumping procedure overlooks the fine structure among the correlations as many joint degrees can contribute to a given overall degree. Indeed it is precisely this structure which acts as a fingerprint of a network ensemble.

Let us introduce the conditional probability $P_{\text{GCC}}(\bm k_{\bm \tau,1}\mid \bm k_{\bm \tau,0})$ that the nearest neighbour has joint degree $\bm k_{\bm \tau,1}$ given that the focal vertex has joint degree $\bm k_{\bm \tau,0}$ in the GCC. Applying Bayes' theorem to our discrete multivariate joint probability we have 
\begin{equation}
    P_{\text{GCC}}(k_{\bot,1},\dots,k_{\gamma,1}\mid k_{\bot,0},\dots,k_{\gamma,0}) = \frac{P_{\text{GCC}}(k_{\bot,0},\dots,k_{\gamma,0}\mid k_{\bot,1},\dots,k_{\gamma,1})P_{\text{GCC}}(k_{\bot,1},\dots,k_{\gamma,1})}{\sum\limits_{k_{\bot,1},\dots,k_{\gamma,1}}P_{\text{GCC}}(k_{\bot,0},\dots,k_{\gamma,0}\mid k_{\bot,1},\dots,k_{\gamma,1})P_{\text{GCC}}(k_{\bot,1},\dots,k_{\gamma,1})}
\end{equation}
Which simplifies to
\begin{equation}
    P_{\text{GCC}}(\bm k_{\bm \tau,1}\mid \bm k_{\bm \tau,0}) = \frac{P_{\text{GCC}}(\bm k_{\bm \tau,0},\bm k_{\bm \tau,1})}{\sum\limits_{\bm k_{\bm \tau,1}}P_{\text{GCC}}(\bm k_{\bm \tau,0},\bm k_{\bm \tau,1})}
\end{equation}
Inserting Eq \ref{eq:central} we find
\begin{equation}
    P_{\text{GCC}}(\bm k_{\bm \tau,1}\mid \bm k_{\bm \tau,0}) = \frac{\sum\limits_{\tau\in \bm \tau}m_{\tau}p_{\bm{k}_{\bm{\tau},0}}{{k}_{{\tau},0}}q_{\tau,{\bm{k}_{\bm{\tau},1}}}\bigg(1-u_\tau^{m_\tau(k_{\tau,0}+k_{\tau,1}-1)-1}\prod\limits_{\nu\in \bm \tau\backslash \tau}u_\nu^{m_\nu(k_{\nu,0} + k_{\nu,1})}\bigg)}{\sum\limits_{\tau\in \bm \tau}\sum\limits_{k_{\tau,1}}m_{\tau}p_{\bm{k}_{\bm{\tau},0}}{{k}_{{\tau},0}}q_{\tau,{\bm{k}_{\bm{\tau},1}}}\bigg(1-u_\tau^{m_\tau(k_{\tau,0}+k_{\tau,1}-1)-1}\prod\limits_{\nu\in \bm \tau\backslash \tau}u_\nu^{m_\nu(k_{\nu,0} + k_{\nu,1})}\bigg)}\label{eq:conditional}
\end{equation}
\begin{figure}
\begin{center}
\includegraphics[width=1.\textwidth]{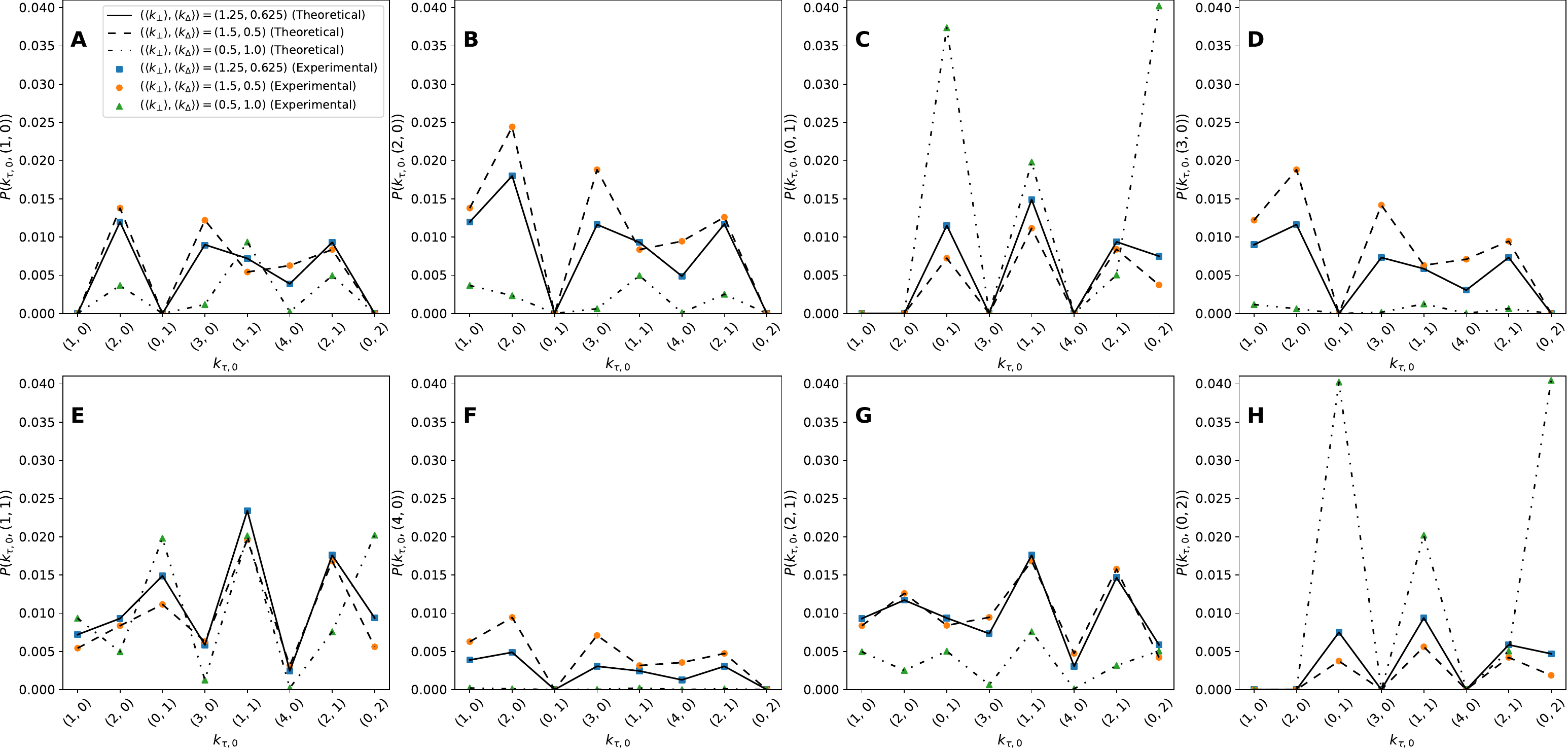}
\caption[cliquePKKmixed]{ 
The probability $P_{\text{GCC}}(s_0,t_0,s_1,t_1)$ for Poisson random graphs comprising of mixed 2-clique and 3-clique topologies for three different clustering regimes. In each plot, the joint degrees of the focal vertex up to overall degree $k=4$ are plotted on the horizontal axis for a given $(s_1,t_1)$ neighbour. Scatter points are the average of 250 repetitions of Monte Carlo simulation on networks with $2\times 10^{5}$ vertices; whilst lines are the analytical results of Eq \ref{eq:central}. The legend is the same as tile (A) for all plots. 
} \label{fig:cliquemixed}
\end{center}
\end{figure}
\end{widetext}

We can use $P_{\text{GCC}}(\bm k_{\bm \tau,1}\mid \bm k_{\bm \tau,0})$ to find multivariate conditional expectation values for a given focal vertex joint degree, generalising \cite{PhysRevLett.87.258701} for the GCM. The expectation value for vector $\bm X$ given vector $\bm Y$ is a vector $\mathcal E[\bm {X}\mid \bm Y] = (\mathcal E [X_1\mid Y],\dots,\mathcal E [X_n\mid \bm Y])^{\text{T}}$ whose elements are the expected values of each of the variables defined as 
\begin{equation}
    \mathcal E[X_i\mid \bm Y=\bm y] = \sum\limits_{x_1,\dots,x_n}x_i P_{\text{GCC}}(x_1,\dots,x_n \mid \bm Y = \bm y)
\end{equation}
For instance, the average joint degree of a neighbour to a focal vertex whose joint degree is $\bm {k_{\tau,0}}$ is the vector $(\mathcal E [ {k_{\bot,1}}\mid \bm {k_{\tau,0}}], \dots, \mathcal E [ {k_{\gamma,1}}\mid \bm {k_{\tau,0}}])^T$ whose elements are 
\begin{align}
     \mathcal E[k_{\tau,1}\mid \bm k_{\bm \tau,0}] =&\ \sum\limits_{ \bm k_{\bm \tau,1}}k_{\tau,1} P(\bm k_{\bm \tau,1} \mid \bm k_{\bm \tau,0})\label{eq:average_deg}
\end{align}
We examine this expression in Appendix \ref{sec:appendixA} for the tree-triangle model. 

\section{Discussion}
\label{sec:discussion}

In this paper we have introduced a theoretical model, based on generating functions, to investigate the NNDC in the GCC of random clustered graphs, constructed according to the GCM, comprising of higher-order clique clusters. We now examine a series of pertinent examples of this model.
\subsection{Single topology }
In the special case that the network consists of a single homogeneous subgraph (a homogeneous subgraph is one where all vertices are degree-equivalent), then $P_{\text{GCC}}(k_{\tau,0},k_{\tau,1})$ from Eq \ref{eq:central} is given by
\begin{equation}
    P_{\text{GCC}} = \frac{(1-u_\tau^{m_\tau(k_{\tau,0}+k_{\tau,1} -1)-1})}{1-u_\tau^{m_\tau\omega_\tau}}q_{\tau,k_{\tau,0}}q_{\tau,k_{\tau,1}}\label{eq:singletopologymain}
\end{equation}
and similarly from Eq \ref{eq:conditional} we have the related conditional probability
\begin{equation}
    P_{\text{GCC}}(k_{\tau,1}\mid k_{\tau,0}) =  \frac{(1-u_\tau^{m_\tau(k_{\tau,0}+k_{\tau,1} -1)-1})}{1-u_\tau^{m_\tau k_{\tau,0}}}q_{\tau,k_{\tau,1}}
\end{equation}
which reproduces the results of \cite{PhysRevE.77.036124,PhysRevE.97.042318} for the nearest-neighbour distributions on the GCC of tree-like networks when $\tau=\bot$. We examine the NNDCs for single-topology networks with Poisson distribution participation in motifs with fixed overall mean degree $\langle k\rangle = 2.5$ in Fig \ref{fig:cliquePKKsingle}. The networks are composed of discrete clique topologies; specifically 2, 3 and 4-cliques in Fig \ref{fig:cliquePKKsingle} A, B and C, respectively. The markers are the averaged results of Monte Carlo simulation while the lines are the theoretical predictions of Eq \ref{eq:singletopologymain}; both are in excellent agreement. In each case, $P_{\text{GCC}}(k_{\tau,0},k_{\tau,1})$ is plotted as a function of increasing $k_{\tau,0}$ for several $k_{\tau,1}$ values. We note that for each clique size $P_{\text{GCC}}(1,1) = 0$; since, this combination cannot exist in the GCC. 
For networks comprised of a single topology, the average degree of a neighbour can be found from Eq \ref{eq:average_deg} as
\begin{equation}
    \mathcal E[ k_{\tau,1}\mid k_{\tau,0}] =  \frac{\sum\limits_{k_{\tau,1}}k_{\tau,1}q_{\tau,k_{\tau,1}}(1-u_\tau^{m_\tau(k_{\tau,0}+k_{\tau,1} -1)-1})}{1-u_\tau^{m_\tau k_{\tau,0}}}\label{eq:ave_degree}
\end{equation}
which is in agreement with \cite{Mizutaka_2020} for tree-like topologies.

\subsection{Tree-triangle model }
We now examine how clustering influences the degree correlations in the GCC of the mixed topology tree-triangle model. The theoretical details of this model are derived in Appendix \ref{sec:appendixA}. Fixing the first moment of the model to $\langle k \rangle=2.5$ the limiting cases of $\langle k_\bot\rangle=0$ and $\langle k_\Delta\rangle=0$ are presented in Fig \ref{fig:cliquePKKsingle} and we now examine i) an even neighbour distribution by setting $\langle k_\bot\rangle =1.25$ and $\langle k_\Delta\rangle = 0.625$; ii) a weakly clustered regime with $\langle k_\bot\rangle =1.5$ and $\langle k_\Delta\rangle = 0.5$ and finally iii) a strong clustering regime with $\langle k_\bot\rangle =0.5$ and $\langle k_\Delta\rangle = 1.0$ in Fig \ref{fig:cliquemixed}. The joint degree of the horizontal axis is ordered by increasing overall degree. When a given overall degree can be formed in multiple ways, such as $k=2$ from $(2,0)$ or $(0,1)$, the degenerate cases are ordered by increasing local clustering coefficient. Each tile in Fig \ref{fig:cliquemixed} A-H plots a given neighbour joint degree (as a function of the focal vertex joint degree) for the three clustering regimes. We observe some encouraging results from these plots: firstly, as with the results of experiments with single-topology networks (Fig \ref{fig:cliquePKKsingle}), the probabilities $P_{\text{GCC}}(1,0,1,0)$ and $P_{\text{GCC}}(0,1,0,1)$ are both zero for the vertices in the GCC (see Fig \ref{fig:cliquemixed} A). We also notice that $P_{\text{GCC}}(s_0,t_0,s_1,t_1)$ takes zero values for impossible combinations, such as neighbours whose edges are of a single, yet opposite, topology to one another. Further, the probabilities are symmetric such that $P_{\text{GCC}}(k_{\bm \tau,0},k_{\bm \tau,1}) = P_{\text{GCC}}(k_{\bm \tau,1},k_{\bm \tau,0})$ which is an expected result for undirected random graphs. Among the non-zero combinations we observe that some peaks, particularly among focal vertices with non-zero degrees in both topologies, are aligned across all series; for example $P_{\text{GCC}}(1,1,1,1)$ in E. Conversely, other peaks such as $P_{\text{GCC}}(2,0,2,1)$ in G peak in the weak and even regimes, yet trough in the strong clustered regime.

We also observe, across all tiles in Fig \ref{fig:cliquemixed} that the correlations among the weak (blue squares) and even-neighbour (orange circles) regimes are generally of higher magnitude across all focal vertices than the strongly clustered regime (green triangles). In other words, the networks with strong clustering exhibit NNDC that have smaller magnitudes with the exception of tiles C and H, which consider neighbouring vertices that only have triangle motifs. 

In tile F we notice that vertices with a high tree-like degree do not tend to connect with neighbours with triangles, especially in the strong clustering regime. 

Collectively, these results give insight into how the network is held together at the microscopic level and how the presence of clustering alters this structure. This could prove useful for creating synthetic networks or for a better understanding of network resilience under targeted attack.

\subsection{The effect of clique size on NNDC}

\begin{figure}
\begin{center}
\includegraphics[width=0.45\textwidth]{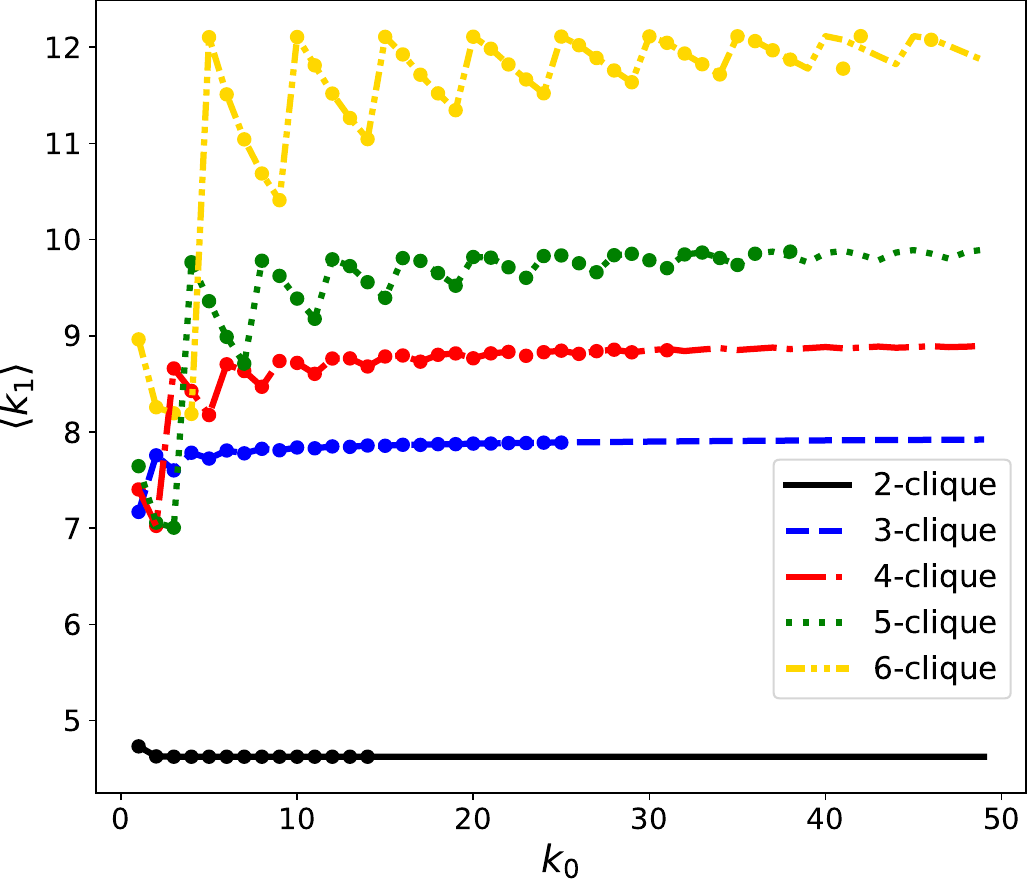}
\caption[h]{ 
The average overall degree of a neighbour for increasing focal vertex degree for binary-topology networks comprising 2-cliques and higher-order cliques. Scatter points are the average of 1000 repetitions of Monte Carlo simulation whilst the plotted lines are the result Eq \ref{Eq:binarymodel}, collected by overall degree according to Eq \ref{eq:lumpeddegrees}. The networks are created from the GCM algorithm with Poisson marginal distributions of each motif topology and overall average degree fixed at $\langle k\rangle=6$ with $\langle k_\bot\rangle =1.25$ across all experiments.
} \label{fig:higherorder}
\end{center}
\end{figure}

In this section, we examine the effect of increasing the clique size on the NNDC of mixed topology GCM networks. To achieve this, we extend the calculations performed in appendix \ref{sec:appendixA} from the 2- and 3-clique model to a binary model composed of 2- and $m$-cliques, whose topology we denote by $\sigma$. For this model, the NNDC for a focal vertex with $s_0$ ordinary edges and $c_0$ edge-disjoint $m$-cliques in the GCC of a GCM network can be obtained from
\begin{widetext}
\begin{align}
   P_{\text{GCC}}(s_0,c_0, s',c')=\frac{p_{s_0c_0}s_0q_{\bot,(s',c')}\left(1-u_\bot^{s_0+s'-2}u_\sigma^{m_\sigma(c_0+c')}\right)+m_\sigma c_0p_{s_0c_0}q_{\sigma,(s',c')}\left(1-u_\bot^{s_0+s'} u_\sigma^{m_\sigma(c_0+c'-1)-1}\right)}{\langle s\rangle(1-u_\bot^2)+m_\sigma\langle c\rangle(1-u_\sigma^{\omega_\sigma})}\label{Eq:binarymodel}
\end{align}
\end{widetext}
The results of this expression are shown in Fig \ref{fig:higherorder}, where the overall neighbour degree is plotted against the overall degree of the focal vertex for several increasing clique sizes. The scatter points are the results of Monte Carlo simulation of networks with 100000 vertices, whilst the plotted lines are the theoretical results of the model; both show excellent agreement with one another. The networks are constructed according to the GCM algorithm before the GCC is selected from the possibly disconnected graph. The motifs counts at each vertex are drawn from Poisson distributions with averages chosen such that the first moment of the distribution of overall degrees is fixed at $\langle k\rangle=6$ across all experiments whilst the average 2-clique count is held fixed at $\langle k_\bot\rangle=1.25$ and the average clique count $\langle k_\sigma\rangle$ is the solution of $\langle k\rangle = \langle k_\bot\rangle +m_\sigma\langle k_\sigma\rangle$. From Fig \ref{fig:higherorder} we observe that the average neighbour degree of networks with larger cliques increases. For cliques larger than 2-cliques, oscillations in the average neighbour degree appear at low focal vertex degree. The amplitude of the oscillations increases with clique size. In each case, the oscillations dampen to a fixed value in the limit of large focal vertex degree.

\subsection{Emergence of correlations}
At criticality, as the GCC emerges, we have that $u_\tau\rightarrow 1$; the probability of not belonging to the GCC is near unity. In this case, the multivariate limit of Eq \ref{eq:central} does not exist. However, in the case that the network is composed of cliques of various sizes which are each independently Poisson distributed at each vertex such that 
\begin{equation}
    p_{\bm k_{\bm\tau,l}} = q_{\tau,\bm k_{\bm\tau,l}} =  \prod_{\tau\in \bm\tau}e^{-\langle k_{\tau}\rangle}\frac{\langle k_{\tau}\rangle^{k_{\tau,l}}}{k_{\tau,l}!}\qquad \forall \tau\in \bm \tau\label{eq:Pois}
\end{equation}
we have that $u_\tau = u^{m_\tau}, \forall \tau$ \cite{karrer_newman_2010}. In this instance Eq \ref{eq:central} is a univariate distribution and we can use L'H\^{o}pital's rule to determine the expected limit to be
\begin{align}
    \lim\limits_{u\rightarrow 1}P_{\text{GCC}}(k_{\tau,0},k_{\tau,1}) = \frac{\sum\limits_{\tau} m_\tau p_{\bm k_{\bm \tau,0}}k_{\tau,0}\Lambda_\tau q_{\tau,\bm k_{\bm \tau,1}}}{\sum\limits_\tau m_\tau^2\omega_\tau \langle k_\tau\rangle }
\end{align}
where 
\begin{equation}
    \Lambda_\tau=m_\tau(k_{\tau,0}+k_{\tau,1}-1)-1 + \sum\limits_{\nu\neq \tau}m_\nu(k_{\nu,0}+k_{\nu,1})
\end{equation}
The critical point can be found by linearising $u_\tau=G_{1,\tau}(\bm {u_\tau^{m_\tau}})$ in a small perturbation $\epsilon$ around $u_\tau = 1- \epsilon_\tau$ \cite{PhysRevE.103.012313}. To leading order in the small parameter $\epsilon_\tau$ we have $\bm \epsilon = \bm A \bm \epsilon$ with $\bm \epsilon = [\epsilon_\bot,\epsilon_\Delta,\dots]^{\text{T}}$. The GCC forms at the point when the determinant $\det|A-I|$ vanishes, where $A =[\partial \bm {\text{G}} /\partial u_\tau]$, $\bm {\text{G}} = [G_{1,\tau},G_{1,\Delta},\dots,G_{1,\gamma}]$ and identity matrix $I$. With mixed topology networks a GCC can form in many different ways. For instance, the GCC of a random graph model with two topologies can form by three distinct mechanisms: a GCC can emerge solely in either of the topologies or global connectivity can occur through a mixture of the binary topologies. 

As we approach the critical point from below, we introduce a characteristic scale $\kappa_\tau$ \cite{song_havlin_makse_2005} associated to the joint degrees of the focal vertex and a neighbour given by $ u_\tau = e^{-1/\kappa_\tau}$. Inserting this expression into Eq \ref{eq:central} for finite $\kappa_\tau$ in each topology, the correlations fall exponentially with increasing $\kappa_\tau$ and hence $P_{\text{GCC}}(k_{\tau,0},k_{\tau,1})$ tends to the uncorrelated value of 
\begin{equation}
    \sum_{\tau\in \bm \tau}m_{\tau}p_{\bm{k}_{\bm{\tau},0}}{{k}_{{\tau},0}}q_{\tau,{\bm{k}_{\bm{\tau},1}}}/\sum_{\tau\in \bm \tau}m_\tau \langle k_\tau\rangle
\end{equation}
Therefore, when the joint degree exceeds the characteristic scale, the GCC is uncorrelated. It is clear that as $u_\tau$ approaches unity the scale diverges $\kappa_\tau \rightarrow \infty$ and hence, the GCC always exhibits degree correlations. In addition, approaching the critical point, the average joint degree (Eq \ref{eq:ave_degree}) falls exponentially with increasing degree along each topology for fixed $\kappa_\tau$. 
\begin{equation}
    \mathcal E[ k_{\tau,1}\mid k_{\tau,0}] =  \frac{\sum\limits_{k_{\tau,1}}k_{\tau,1}q_{\tau,k_{\tau,1}}(1-e^{-\phi})}{1-e^{-m_\tau k_{\tau,0}/\kappa_\tau}}
\end{equation}
where $\phi = m_\tau(k_{\tau,0}+k_{\tau,1} -1)-1/\kappa_\tau$. Thus, the correlations which are present at the critical point are negative in nature. It might happen, however, given the number of ways that the GCC of a mixed motif random graph model can emerge, that the characteristic scales of all topologies don't diverge at the critical point. For instance, consider a doubly Poisson distributed tree-triangle model with a critical average tree degree, but a sub-critical average triangle degree. A GCC will form among the tree edges, but the probability of those vertices involved only in triangles, $(0,t)$ for $t=1,2,3,\dots$, connecting to this GCC is small; since, their connection requires them to connect to mixed-topology vertices, which in turn connect to the GCC. Thus, we might find that the negative degree correlation structure among the triangles has not yet formed despite there being a non-zero density of triangles in the GCC.

\subsection{Empirical networks}
\label{sec:empirical}

We now examine the correlation properties of the GCC of the ensemble representation of empirical networks using our joint degree model. Random graphs are elements of an ensemble $\mathcal G$ of graphs with $V$ vertices and $E$ edges; each member occurring with probability $P(G)$ \cite{PhysRevE.77.036124}. The average value of a property of graph $G$, $Z(G)$, (such as its degree distribution or average degree) can be averaged over the entire ensemble
\begin{equation}
    \langle Z\rangle = \sum_{G\in \mathcal G}Z(G)P(G)
\end{equation}
The generating function formulation describes the properties of the ensemble. Empirical networks $g$ are particular realisations of members of $\mathcal G$. The properties of a particular realisation are given by 
\begin{equation}
    P(Z) = \sum_{G\in \mathcal G}\delta(Z-Z(G))P(G)
\end{equation}
If $P(Z)$ is well represented by the ensemble average then the generating function formulation can be used to describe the properties of $g$. To study the NNDC in the GCC of $g$ using generating functions, we must represent the largest component of an empirical network by a joint degree sequence of subgraphs. Whilst the choice of subgraphs is arbitrary \cite{PhysRevE.103.012309}, we only include cliques in the topology representation due to the vast literature on clique finding algorithms and the simplicity of calculating their properties. The clique decomposition of the GCC of $g$ whose cliques have order less than or equal to $\omega$ can be performed in many different ways; and the resulting joint degree sequence can exhibit significantly different properties in terms of the number of subgraphs present their clustering, and other properties. Given that the method to create the joint degree distribution is not unique, and that the ensemble properties of each particular decomposition are often dissimilar, we now examine three clique decompositions and compare their properties.  

The trivial decomposition is to simply cover $g$ with 2-cliques; we refer to this as the single-edge-decomposition (SED). The degree sequence can then be used to create realisations using the ordinary configuration model. Another simple cover is the minimal cover of maximal cliques. However, it is very likely that the edges of the cliques will not be disjoint, i.e. a single edge will be a member of more than one clique. Whilst this could be an accurate representation of a vertex's local environment, the construction process for random graphs using the GCM will not work. Thus, we must impose that the cover is edge-disjoint.  

\begin{figure}
\begin{center}
\includegraphics[width=0.46\textwidth]{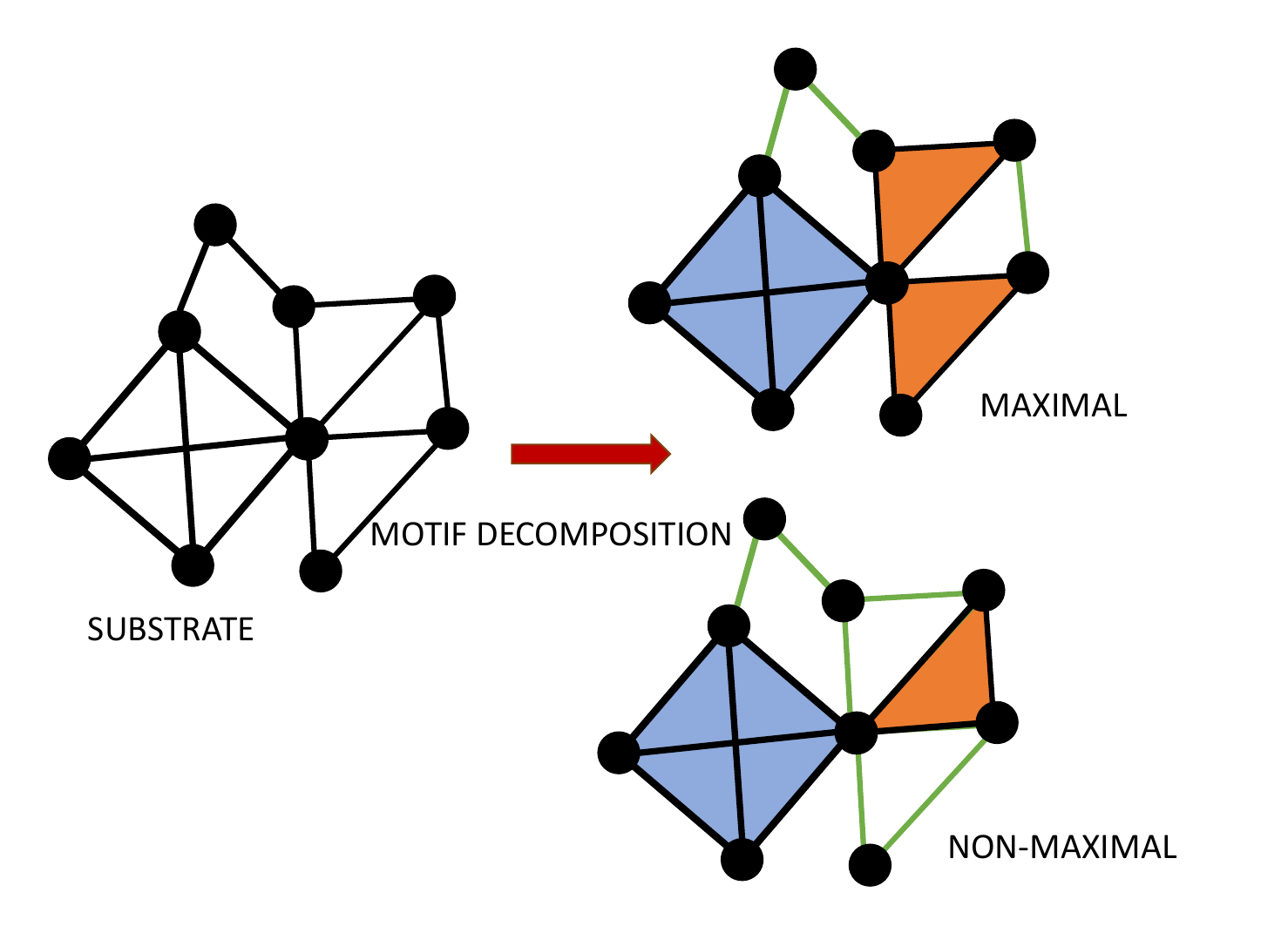}
\caption[motifs]{ 
The clique decomposition of a substrate network (left) can be performed in multiple ways. Two examples are shown (right). The shaded faces are higher-order cliques whilst the green edges are 2-cliques. The clustering of the resulting joint degree distributions (and their random graph ensembles) are significantly altered depending on how the decomposition is performed. The maximal representation has 6 cliques in total whilst the non-maximal representation has 8 cliques. When only maximal representations are extracted the decomposition is a EECC. 
} \label{fig:motifs}
\end{center}
\end{figure}

One proposed method of clique decomposition is defined heuristically as follows \cite{2021arXiv210103618B}: we obtain the set $C$ of all maximal cliques from the network; each maximal $n$-clique $c_i\in C$, $n\in \{1,\dots, \omega\}$ is scored according to the fraction of edges it shares with other members of $C$. The largest clique within the set of lowest score cliques are included in the representation and $C$ is recalculated. The process is repeated until the edges of the substrate network are expended. Such a covering is known as a edge-disjoint edge clique cover (EECC), see Fig \ref{fig:motifs} for details.

\begin{figure}
\begin{center}
\includegraphics[width=0.35\textwidth]{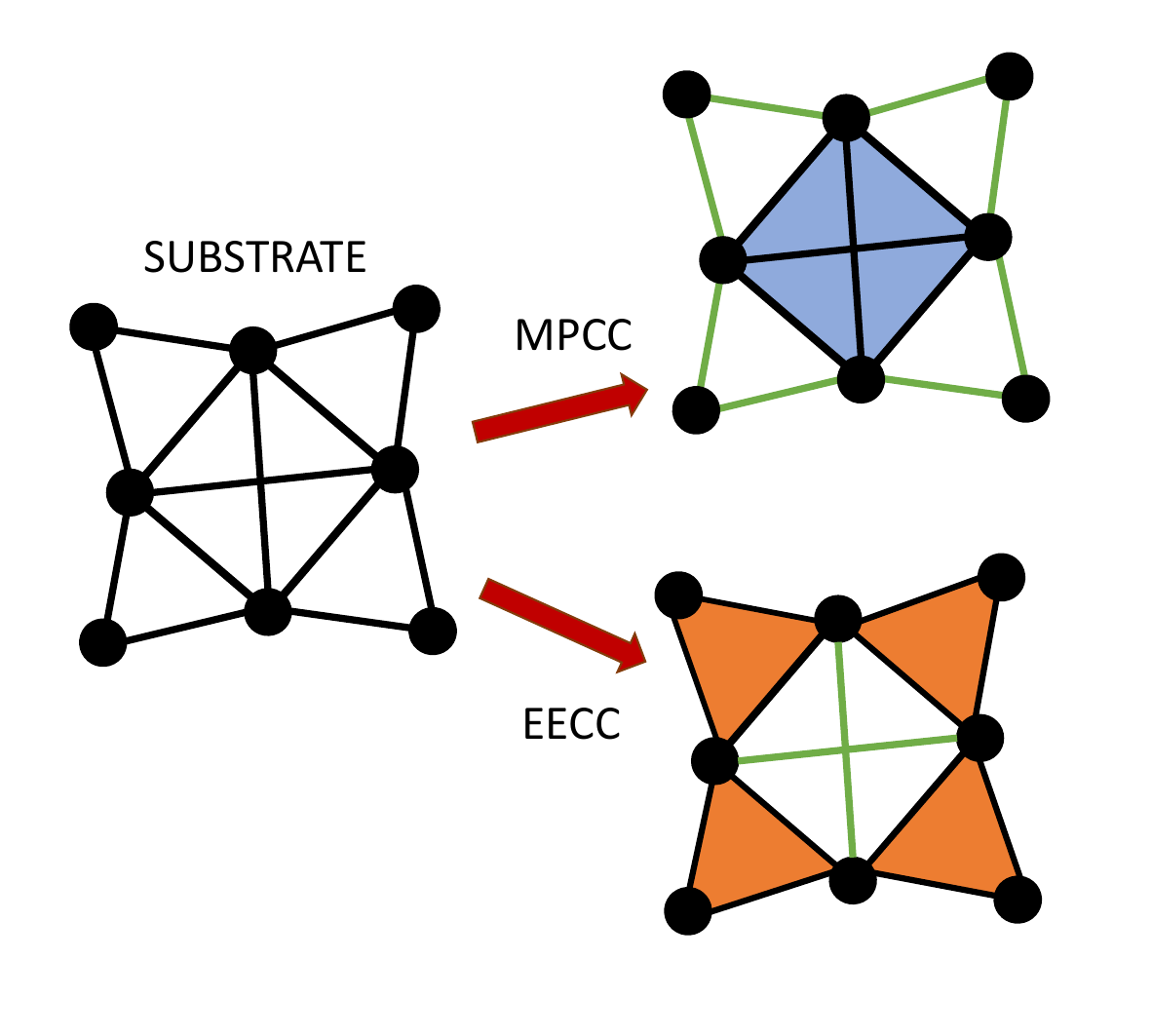}
\caption[motifs2]{ 
The results of the to clique decomposition algorithms (MPCC) and (EECC) for a particular substrate graph. The MPCC favours the formation of large subgraphs, leading to 9 cliques (a single 4-clique and 8 2-cliques) whilst the EECC leads to 6 cliques (4 3-cliques and 2 2-cliques). The joint degree sequence obtained from the MPCC network creates a non-maximal random ensemble of GCM networks.
} \label{fig:motifs2}
\end{center}
\end{figure}

\begin{figure}
\begin{center}
\includegraphics[width=0.35\textwidth]{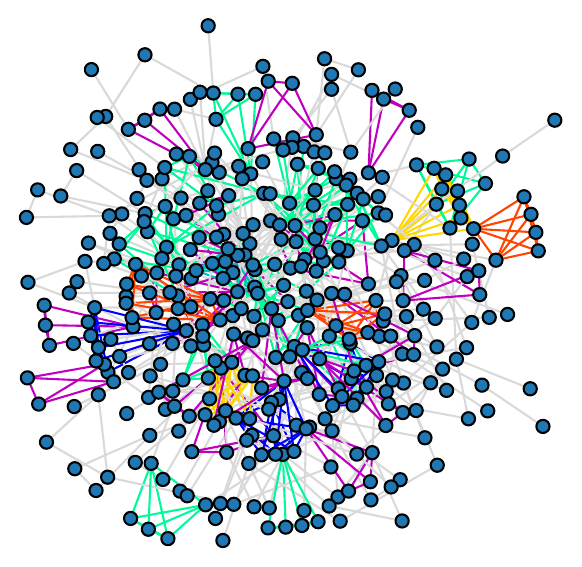}
\caption[netsci]{ 
A member of the MPCC random graph ensemble of the GCC of the network science authorship network with higher-order cliques (larger than 3-cliques) coloured for clarity. Specifically, the 4-cliques are magenta, 5-cliques are light green, 6-cliques are orange, 7-cliques are blue, 8-cliques are yellow and the 9-clique is cyan. Unlike random graphs constructed using the EECC method, larger cliques are preferentially retained in the ensemble.
} \label{fig:netsci}
\end{center}
\end{figure}

\begin{figure}
\begin{center}
\includegraphics[width=0.475\textwidth]{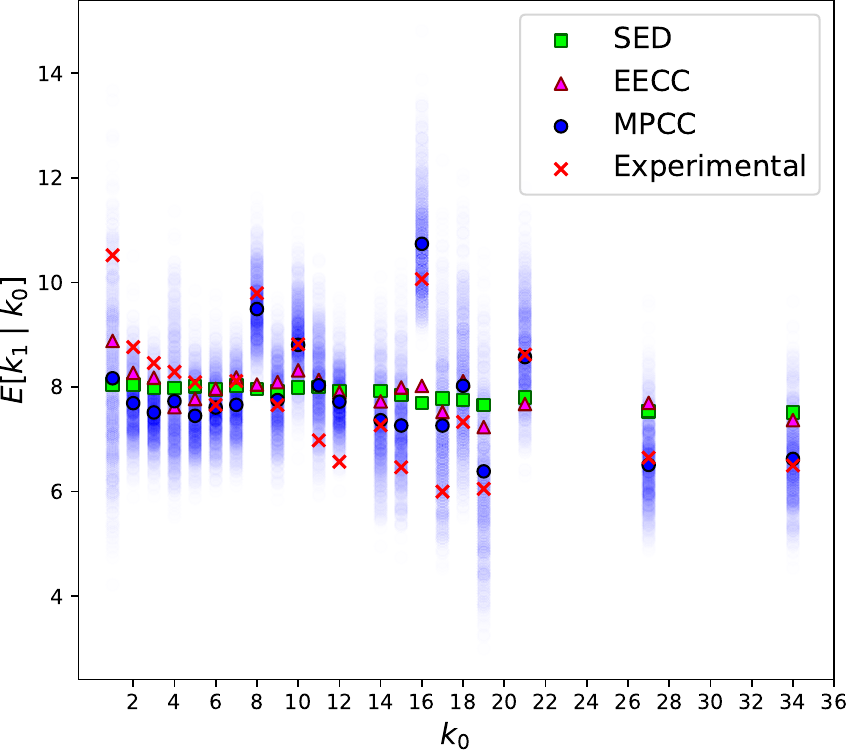}
\caption[netsci]{ 
The ensemble expectation value of the  overall degree of a neighbour as a function of focal vertex degree for clique covers of the network science authorship network. Plotted are the experimental results (red crosses), the average EECC (pink triangles), the average MPCC (dark blue circles) and its variance (light blue circles) for each realisation. Each simulation was performed 1000 times. The SED (green squares) doesn't capture the correlation structure for this network. The MPCC accurately captures the correlation structure of the high-degree vertices due to retaining the larger motifs that a vertex belongs to; however, the low (mid) degree sites are generally under (over) predicted. Conversely, the EECC performs well for the low and mid-degree vertices, but tends to the SED for the high-degree sites. 
} \label{fig:netsciexp} 
\end{center}
\end{figure}

We propose a novel alternative clique cover as follows: the set $\mathcal C$ of all cliques present in the network (including those induced from subgraphs of larger cliques) is obtained from the empirical network. The set is ordered such that the largest cliques have the highest precedence. The subset of cliques within $\mathcal C$ that have equal size $\forall n\in \{1,\dots,\omega\}$ are then scored in a similar fashion to the EECC algorithm and the cliques with the lowest score (and therefore the least number of overlapping edges with other motifs) are given highest precedence. The order of cliques with equivalent size and score is then randomised, thus the cover is stochastic. The largest cliques are drawn from $\mathcal C$ and placed on the network if their edges do not overlap other with cliques that have already been placed in the network. The list is iterated until all edges belong to an independent clique. This method draws non-maximal joint degree sequences; however, higher-order cliques are preferentially preserved, we describe it as an edge disjoint motif preserving edge clique cover (MPCC), see Fig \ref{fig:motifs2}. In the particular case that the set of maximal cliques are edge disjoint, the distribution obtained from both the EECC and MPCC motif decomposition algorithms are in agreement with one another. It should be mentioned that both covers are not unique when two cliques of a given size and score can be chosen. Within the MPCC, we 
resolve these degeneracies by retaining the cliques associated with higher degree vertices. In our implementation of the EECC, we choose cliques from the set of degenerate cliques at random.
Once a suitable cover has been formed for the network, its joint-degree sequence can be extracted. This sequence is then used to create an ensemble of GCM networks. As a concrete example of this method we extract the joint degree sequences, using the SED, EECC and the MPCC, of the GCC of the network science authorship network \cite{PhysRevE.74.036104} and use the GCM algorithm to construct random graph ensembles, Fig \ref{fig:netsci}. Plotted in Fig \ref{fig:netsciexp} are the experimental results from the original network (red crosses), the SED (green squares), the EECC (pink triangles) and the results from the MPCC algorithm (light blue circles) as well as their average (dark blue circles). The average neighbour degree, $k_1$ obtained from the SED shows poor accuracy when compared to the experimental results. Instead of the detailed NNDC structure over the range of focal vertex degrees, the neighbour degrees tend to fluctuate around $k_1=8$. In contrast, the MPCC exhibits a rich correlation structure whose average follows the trends of the experimental data. Additionally, the average neighbour degree for the high-degree vertices is well represented; however, this is at the expense of the lower degree information, where the representation is less accurate. The EECC shows fair agreement across the range of focal vertex degrees, outperforming the MPCC at low degrees; however, the MPCC represents the empirical network correlations for the high-degree vertices with greater accuracy than the EECC. The EECC representation of the high-degree sites is in agreement with the SED, indicating that these cliques are destroyed during the covering process. We notice from the variance of the MPCC that the NNDC of the empirical network is dense within the set of ensemble representations. 

\section{Conclusion}
\label{sec:conclusion}

In this paper we have introduced a robust analytical framework to study the NNDC between vertices in the GCC of random graphs constructed according to the GCM. We have used our method to investigate the correlation properties of synthetic clustered GCM graphs in detail and found they exhibit organisation among their subgraphs. We studied the behaviour of the NNDC as the size of the substrate motif increases, along with the clustering for a fixed first moment of the overall average degree. We found that the NNDC among networks composed from  larger cliques tend to be larger in magnitude for low degree vertices due to the constraint on the first moment of the overall degree. 

 Investigating the tree-triangle model in detail, we found that the joint degrees are negatively correlated along each topology as found for tree-like topologies in other studies \cite{PhysRevE.77.036124,PhysRevE.97.042318,PhysRevE.98.062314}.
 

The magnitude and the patterns of NNDC were found to vary significantly with the clustering coefficient of the network ensemble. The correlations among neighbours of mixed topology focal vertices in tree-triangle networks with larger clustering coefficients were smaller in magnitude, in general, with respect to the single-topology vertices.  

We then investigated the role of clique size for GCM graphs and observed oscillations in the average overall neighbour degrees as a function of focal vertex degree. We found that the average neighbour degree in the GCC increases for networks composed of larger cliques.

Lastly, we studied the correlation structure of the random graph ensemble of an empirical network. To do this, we introduced a novel clique decomposition algorithm and compared it to other heuristics in the literature. We found that the manner in which the network is decomposed into motifs greatly effects the correlation substructure of the ensemble representation. 

This work increases our understanding of the NNDC of clustered networks comprised of higher-order clique motifs; however, we have not addressed the long range correlation structure or defined an assortativity coefficient for these graphs, which we leave for future work.

\section{ACKNOWLEDGMENTS}

This work was partially supported by the UK Engineering and Physical Sciences Research Council under grant number EP/N007565/1 (Science of Sensor Systems Software).


\bibliography{bib}

@article{Mizutaka_2020,
	doi = {10.1088/2632-072x/abb4c5},
	url = {https://doi.org/10.1088/2632-072x/abb4c5},
	year = 2020,
	month = {sep},
	publisher = {{IOP} Publishing},
	volume = {1},
	number = {3},
	pages = {035007},
	author = {Shogo Mizutaka and Takehisa Hasegawa},
	title = {Emergence of long-range correlations in random networks},
	journal = {Journal of Physics: Complexity}
}

@article{PhysRevE.98.062314,
  title = {Disassortativity of percolating clusters in random networks},
  author = {Mizutaka, Shogo and Hasegawa, Takehisa},
  journal = {Phys. Rev. E},
  volume = {98},
  issue = {6},
  pages = {062314},
  numpages = {7},
  year = {2018},
  month = {Dec},
  publisher = {American Physical Society},
  doi = {10.1103/PhysRevE.98.062314},
  url = {https://link.aps.org/doi/10.1103/PhysRevE.98.062314}
}

@article{PhysRevE.101.062310,
  title = {Structure of percolating clusters in random clustered networks},
  author = {Hasegawa, Takehisa and Mizutaka, Shogo},
  journal = {Phys. Rev. E},
  volume = {101},
  issue = {6},
  pages = {062310},
  numpages = {12},
  year = {2020},
  month = {Jun},
  publisher = {American Physical Society},
  doi = {10.1103/PhysRevE.101.062310},
  url = {https://link.aps.org/doi/10.1103/PhysRevE.101.062310}
}

@book{newman_2019, 
 place={Oxford}, 
 title={Networks}, 
 publisher={Oxford University Press}, author={Newman, Mark E.J}, 
 year={2019}}

@article{PhysRevE.83.036112,
  title = {The unreasonable effectiveness of tree-based theory for networks with clustering},
  author = {Melnik, Sergey and Hackett, Adam and Porter, Mason A. and Mucha, Peter J. and Gleeson, James P.},
  journal = {Phys. Rev. E},
  volume = {83},
  issue = {3},
  pages = {036112},
  numpages = {12},
  year = {2011},
  month = {Mar},
  publisher = {American Physical Society},
  doi = {10.1103/PhysRevE.83.036112},
  url = {https://link.aps.org/doi/10.1103/PhysRevE.83.036112}
}

@article{PhysRevE.97.062308,
  title = {General formulation of long-range degree correlations in complex networks},
  author = {Fujiki, Yuka and Takaguchi, Taro and Yakubo, Kousuke},
  journal = {Phys. Rev. E},
  volume = {97},
  issue = {6},
  pages = {062308},
  numpages = {8},
  year = {2018},
  month = {Jun},
  publisher = {American Physical Society},
  doi = {10.1103/PhysRevE.97.062308},
  url = {https://link.aps.org/doi/10.1103/PhysRevE.97.062308}
}

@article{fosdick_larremore_nishimura_ugander_2018, title={Configuring random graph models with fixed degree sequences}, volume={60}, DOI={10.1137/16m1087175}, number={2}, journal={SIAM Review}, author={Fosdick, Bailey K. and Larremore, Daniel B. and Nishimura, Joel and Ugander, Johan}, year={2018}, pages={315–355}}

@article{Peron_2018,
	doi = {10.1209/0295-5075/121/68001},
	url = {https://doi.org/10.1209/0295-5075/121/68001},
	year = 2018,
	month = {mar},
	publisher = {{IOP} Publishing},
	volume = {121},
	number = {6},
	pages = {68001},
	author = {Thomas K. DM. Peron and Peng Ji and Jürgen Kurths and Francisco A. Rodrigues},
	title = {Spectra of random networks in the weak clustering regime},
	journal = {{EPL} (Europhysics Letters)}
}

@article{Stegehuis_2021,
	doi = {10.1088/2632-072x/ac35b7},
	url = {https://doi.org/10.1088/2632-072x/ac35b7},
	year = 2021,
	month = {nov},
	publisher = {{IOP} Publishing},
	volume = {2},
	number = {4},
	pages = {045011},
	author = {Clara Stegehuis and Thomas Peron},
	title = {Network processes on clique-networks with high average degree: the limited effect of higher-order structure},
	journal = {Journal of Physics: Complexity}
}

@article{PhysRevE.104.024304,
  title = {Exact formula for bond percolation on cliques},
  author = {Mann, Peter and Smith, V. Anne and Mitchell, John B. O. and Jefferson, Christopher and Dobson, Simon},
  journal = {Phys. Rev. E},
  volume = {104},
  issue = {2},
  pages = {024304},
  numpages = {10},
  year = {2021},
  month = {Aug},
  publisher = {American Physical Society},
  doi = {10.1103/PhysRevE.104.024304},
  url = {https://link.aps.org/doi/10.1103/PhysRevE.104.024304}
}

@article{PhysRevE.74.036104,
  title = {Finding community structure in networks using the eigenvectors of matrices},
  author = {Newman, M. E. J.},
  journal = {Phys. Rev. E},
  volume = {74},
  issue = {3},
  pages = {036104},
  numpages = {19},
  year = {2006},
  month = {Sep},
  publisher = {American Physical Society},
  doi = {10.1103/PhysRevE.74.036104},
  url = {https://link.aps.org/doi/10.1103/PhysRevE.74.036104}
}

@article{PhysRevLett.87.258701,
  title = {Dynamical and Correlation Properties of the Internet},
  author = {Pastor-Satorras, Romualdo and V\'azquez, Alexei and Vespignani, Alessandro},
  journal = {Phys. Rev. Lett.},
  volume = {87},
  issue = {25},
  pages = {258701},
  numpages = {4},
  year = {2001},
  month = {Nov},
  publisher = {American Physical Society},
  doi = {10.1103/PhysRevLett.87.258701},
  url = {https://link.aps.org/doi/10.1103/PhysRevLett.87.258701}
}

@article{2021arXiv210103618B, title={Network clique cover approximation to analyze complex contagions through group interactions}, volume={4}, DOI={10.1038/s42005-021-00618-z}, number={1}, journal={Communications Physics}, author={Burgio, Giulio and Arenas, Alex and G{\'o}mez, Sergio and Matamalas, Joan T.}, year={2021}}

@article{song_havlin_makse_2005, title={Self-similarity of complex networks}, volume={433}, DOI={10.1038/nature03248}, number={7024}, journal={Nature}, author={Song, Chaoming and Havlin, Shlomo and Makse, Hernán A.}, year={2005}, pages={392–395}}

@article{HASEGAWA2021125970,
title = {Observability transitions in clustered networks},
journal = {Physica A: Statistical Mechanics and its Applications},
volume = {573},
pages = {125970},
year = {2021},
issn = {0378-4371},
doi = {https://doi.org/10.1016/j.physa.2021.125970},
url = {https://www.sciencedirect.com/science/article/pii/S0378437121002429},
author = {Takehisa Hasegawa and Yuta Iwase}
}

@article{PhysRevE.103.012309,
  title = {Random graphs with arbitrary clustering and their applications},
  author = {Mann, Peter and Smith, V. Anne and Mitchell, John B. O. and Dobson, Simon},
  journal = {Phys. Rev. E},
  volume = {103},
  issue = {1},
  pages = {012309},
  numpages = {10},
  year = {2021},
  month = {Jan},
  publisher = {American Physical Society},
  doi = {10.1103/PhysRevE.103.012309},
  url = {https://link.aps.org/doi/10.1103/PhysRevE.103.012309}
}

@article{PhysRevE.103.012313,
  title = {Percolation in random graphs with higher-order clustering},
  author = {Mann, Peter and Smith, V. Anne and Mitchell, John B. O. and Dobson, Simon},
  journal = {Phys. Rev. E},
  volume = {103},
  issue = {1},
  pages = {012313},
  numpages = {11},
  year = {2021},
  month = {Jan},
  publisher = {American Physical Society},
  doi = {10.1103/PhysRevE.103.012313},
  url = {https://link.aps.org/doi/10.1103/PhysRevE.103.012313}
}

@article{PhysRevLett.103.058701,
  title = {Random Graphs with Clustering},
  author = {Newman, M. E. J.},
  journal = {Phys. Rev. Lett.},
  volume = {103},
  issue = {5},
  pages = {058701},
  numpages = {4},
  year = {2009},
  month = {Jul},
  publisher = {American Physical Society},
  doi = {10.1103/PhysRevLett.103.058701},
  url = {https://link.aps.org/doi/10.1103/PhysRevLett.103.058701}
}

@article{PhysRevE.80.020901,
  title = {Percolation and epidemics in random clustered networks},
  author = {Miller, Joel C.},
  journal = {Phys. Rev. E},
  volume = {80},
  issue = {2},
  pages = {020901},
  numpages = {4},
  year = {2009},
  month = {Aug},
  publisher = {American Physical Society},
  doi = {10.1103/PhysRevE.80.020901},
  url = {https://link.aps.org/doi/10.1103/PhysRevE.80.020901}
}

@article{PhysRevE.68.026121,
  title = {Properties of highly clustered networks},
  author = {Newman, M. E. J.},
  journal = {Phys. Rev. E},
  volume = {68},
  issue = {2},
  pages = {026121},
  numpages = {6},
  year = {2003},
  month = {Aug},
  publisher = {American Physical Society},
  doi = {10.1103/PhysRevE.68.026121},
  url = {https://link.aps.org/doi/10.1103/PhysRevE.68.026121}
}

@article{miller_2009_spread, title={Spread of infectious disease through clustered populations}, volume={6}, DOI={10.1098/rsif.2008.0524}, number={41}, journal={Journal of The Royal Society Interface}, author={Miller, Joel C.}, year={2009}, pages={1121–1134}}

@article{gleeson_2009, title={Bond percolation on a class of clustered random networks}, volume={80}, DOI={10.1103/physreve.80.036107}, number={3}, journal={Physical Review E}, author={Gleeson, James P.}, year={2009}, month={Oct}}

@article{PhysRevE.77.036124,
  title = {Correlations in connected random graphs},
  author = {Bialas, Piotr and Ole\ifmmode \acute{s}\else \'{s}\fi{}, Andrzej K.},
  journal = {Phys. Rev. E},
  volume = {77},
  issue = {3},
  pages = {036124},
  numpages = {10},
  year = {2008},
  month = {Mar},
  publisher = {American Physical Society},
  doi = {10.1103/PhysRevE.77.036124},
  url = {https://link.aps.org/doi/10.1103/PhysRevE.77.036124}
}

@article{PhysRevE.76.045101,
  title = {Component sizes in networks with arbitrary degree distributions},
  author = {Newman, M. E. J.},
  journal = {Phys. Rev. E},
  volume = {76},
  issue = {4},
  pages = {045101},
  numpages = {4},
  year = {2007},
  month = {Oct},
  publisher = {American Physical Society},
  doi = {10.1103/PhysRevE.76.045101},
  url = {https://link.aps.org/doi/10.1103/PhysRevE.76.045101}
}

@article{Arenas_2008,
	doi = {10.1088/1751-8113/41/22/224001},
	url = {https://doi.org/10.1088/1751-8113/41/22/224001},
	year = 2008,
	month = {may},
	publisher = {{IOP} Publishing},
	volume = {41},
	number = {22},
	pages = {224001},
	author = {A Arenas and A Fern{\'{a}}ndez and S Fortunato and S G{\'{o}}mez},
	title = {Motif-based communities in complex networks},
	journal = {Journal of Physics A: Mathematical and Theoretical}
}

@article{allard_hebert-dufresne_young_dube_2015, title={General and exact approach to percolation on random graphs}, volume={92}, DOI={10.1103/physreve.92.062807}, number={6}, journal={Physical Review E}, author={Allard, Antoine and Hébert-Dufresne, Laurent and Young, Jean-Gabriel and Dubé, Louis J.}, year={2015}, month={Jul}}

@article{PhysRevE.72.036133,
  title = {Tuning clustering in random networks with arbitrary degree distributions},
  author = {\'Angeles Serrano, M. and Bogu\~n\'a, Mari\'an},
  journal = {Phys. Rev. E},
  volume = {72},
  issue = {3},
  pages = {036133},
  numpages = {8},
  year = {2005},
  month = {Sep},
  publisher = {American Physical Society},
  doi = {10.1103/PhysRevE.72.036133},
  url = {https://link.aps.org/doi/10.1103/PhysRevE.72.036133}
}

@article{karrer_newman_2010, title={Random graphs containing arbitrary distributions of subgraphs}, volume={82}, DOI={10.1103/physreve.82.066118}, number={6}, journal={Physical Review E}, author={Karrer, Brian and Newman, M. E. J.}, year={2010}}

@article{PhysRevE.83.056107,
  title = {Cascades on a class of clustered random networks},
  author = {Hackett, Adam and Melnik, Sergey and Gleeson, James P.},
  journal = {Phys. Rev. E},
  volume = {83},
  issue = {5},
  pages = {056107},
  numpages = {9},
  year = {2011},
  month = {May},
  publisher = {American Physical Society},
  doi = {10.1103/PhysRevE.83.056107},
  url = {https://link.aps.org/doi/10.1103/PhysRevE.83.056107}
}

@article{PhysRevE.81.066114,
  title = {How clustering affects the bond percolation threshold in complex networks},
  author = {Gleeson, James P. and Melnik, Sergey and Hackett, Adam},
  journal = {Phys. Rev. E},
  volume = {81},
  issue = {6},
  pages = {066114},
  numpages = {10},
  year = {2010},
  month = {Jun},
  publisher = {American Physical Society},
  doi = {10.1103/PhysRevE.81.066114},
  url = {https://link.aps.org/doi/10.1103/PhysRevE.81.066114}
}

@article{PhysRevE.97.042318,
  title = {Revealing the microstructure of the giant component in random graph ensembles},
  author = {Tishby, Ido and Biham, Ofer and Katzav, Eytan and K\"uhn, Reimer},
  journal = {Phys. Rev. E},
  volume = {97},
  issue = {4},
  pages = {042318},
  numpages = {19},
  year = {2018},
  month = {Apr},
  publisher = {American Physical Society},
  doi = {10.1103/PhysRevE.97.042318},
  url = {https://link.aps.org/doi/10.1103/PhysRevE.97.042318}
}

@article{wang_lizardo_hachen_2014, title={Algorithms for generating large-scale clustered random graphs}, volume={2}, DOI={10.1017/nws.2014.7}, number={3}, journal={Network Science}, author={Wang, Cheng and Lizardo, Omar and Hachen, David}, year={2014}, pages={403–415}}

@article{heath_parikh_2011, title={Generating random graphs with tunable clustering coefficients}, volume={390}, DOI={10.1016/j.physa.2011.06.052}, number={23-24}, journal={Physica A: Statistical Mechanics and its Applications}, author={Heath, Lenwood S. and Parikh, Nidhi}, year={2011}, pages={4577–4587}}

@article{10.1093/comnet/cnw011,
    author = {Ritchie, Martin and Berthouze, Luc and Kiss, Istvan Z.},
    title = "{Generation and analysis of networks with a prescribed degree sequence and subgraph family: higher-order structure matters}",
    journal = {Journal of Complex Networks},
    volume = {5},
    number = {1},
    pages = {1-31},
    year = {2016},
    month = {05},
    issn = {2051-1310},
    doi = {10.1093/comnet/cnw011},
    url = {https://doi.org/10.1093/comnet/cnw011},
    eprint = {https://academic.oup.com/comnet/article-pdf/5/1/1/10253551/cnw011.pdf},
}

@article{PhysRevE.99.042308,
  title = {Generating random networks that consist of a single connected component with a given degree distribution},
  author = {Tishby, Ido and Biham, Ofer and Katzav, Eytan and K\"uhn, Reimer},
  journal = {Phys. Rev. E},
  volume = {99},
  issue = {4},
  pages = {042308},
  numpages = {15},
  year = {2019},
  month = {Apr},
  publisher = {American Physical Society},
  doi = {10.1103/PhysRevE.99.042308},
  url = {https://link.aps.org/doi/10.1103/PhysRevE.99.042308}
}

\appendix
\begin{widetext}
\section{Results within the tree-triangle model}
\label{sec:appendixA}

In this section we derive the expectation values for the tree-triangle model, see Fig \ref{fig:correlations}. For this model the generating function for the number of nearest-neighbours given the joint degree of the focal vertex is $k_{\bm \tau,0} = (s_0,t_0)$ is given by unpacking Eq \ref{eq:FHAT} for $\bm \tau=\{\bot,\Delta\}$. We obtain
\begin{align}
    \hat{F}_{GCC}(\bm x,\bm y , s_0,t_0) =&\ p_{s_0,t_0}f_\bot
 ^{s_0}
  f_\Delta^{2t_0} -
     p_{s_0,t_0}g_\bot^{s_0}
g_\Delta^{2t_0}
\end{align}
where $f_\tau = \sum_s\sum_t q_{\tau,(s,t)}z_{st}$, $g_\bot= \sum_s\sum_t q_{\bot,(s,t)}u_{\bot}^{s-1}u_{\Delta}^{2t}x_{s}y_{t}$ and $\sum_s\sum_t q_{\Delta,(s,t)}u_{\bot}^{s}u_{\Delta}^{2(t-1)}x_{s}y_{t}$. The evaluation of the expectation values for the nearest-neighbours to a vertex of joint degree $(s_0,t_0)$ in the tree-triangle model is given by the following derivative 
\begin{equation}
     \hat{F}_{\text{GCC}}' = \frac{d \hat{F}_{\text{GCC}}}{d z_{s't'}}\bigg|_{z_{s't'}=1}
\end{equation}
We evaluate this as follows
\begin{align}
    \frac{d \hat{F}_{\text{GCC}}}{d z_{s't'}}\bigg|_{z_{s't'}=1} =\ & \frac{d }{d z_{s't'}}\bigg|_{z_{s't'}=1}p_{s_0t_0} f_\bot^{s_0}f_\Delta^{2t_0} -\frac{d }{d z_{s't'}}\bigg|_{z_{s't'}=1}p_{s_0t_0}  g_\bot^{s_0}g_\Delta^{2t_0} 
    \\
    =&\ p_{s_0t_0}\left(s_0f_{\bot}^{s_0-1}\frac{df_\bot}{dz_{s't'}}f_{\Delta}^{2t_0} + 2t_0f_{\bot}^{s_0}f_{\Delta}^{2(t_0-1)}f_\Delta\frac{df_\Delta}{dz_{s't'}}\right)\nonumber\\
    &- p_{s_0t_0}\left(s_0g_{\bot}^{s_0-1}\frac{dg_\bot}{dz_{s't'}}g_{\Delta}^{2t_0} + 2t_0g_{\bot}^{s_0}g_{\Delta}^{2(t_0-1)}g_\Delta\frac{dg_\Delta}{dz_{s't'}}\right)
\end{align}
At $z_{s't'}=1$ we have $f_\tau(1)=1$, $g_\tau(1) = G_{1,\tau}(u_\bot,u_\Delta^2)$ and also
 \begin{align}
     \frac{df_\tau}{dz_{s't'}}\bigg|_{z_{s't'}=1} &= \frac{d}{dz_{s't'}}\sum_s\sum_tq_{\tau,(s,t)}z_{st}\\
     &=q_{\tau,(s',t')}
 \end{align}
 and 
  \begin{align}
     \frac{dg_\bot}{dz_{s't'}}\bigg|_{z_{s't'}=1} &= \frac{d}{dz_{s't'}}\sum_s\sum_tq_{\bot,(s,t)}u_\bot^{s-1}u_\Delta^{2t}z_{st}\\
     &=q_{\bot,(s',t')}u_\bot^{s'-1}u_\Delta^{2t'}
 \end{align}
   \begin{align}
     \frac{dg_\Delta}{dz_{s't'}}\bigg|_{z_{s't'}=1} &= \frac{d}{dz_{s't'}}\sum_s\sum_tq_{\Delta,(s,t)}u_\bot^{s}u_\Delta^{2(t-1)}z_{st}\\
     &=q_{\Delta,(s',t')}u_\bot^{s'}u_\Delta^{2(t'-1)}
 \end{align}
Thus, we find 
\begin{align}
    \frac{d \hat{F}_{\text{GCC}}}{d z_{s't'}}\bigg|_{z_{s't'}=1} =\ &\ p_{s_0t_0}\bigg(s_0q_{\bot,(s',t')} + 2t_0q_{\Delta,(s',t')}\bigg)- p_{s_0t_0}\bigg(s_0u_\bot^{s_0-1}q_{\bot,(s',t')}u_\bot^{s'-1}u_\Delta^{2t'}u_\Delta^{2t_0}\nonumber\\
     &+ 2t_0u_\bot ^{s_0}u_\Delta^{2(t_0-1)}u_\Delta q_{\Delta,(s',t')}u_\bot^{s'}u_\Delta^{2(t'-1)}
     \bigg)\label{eq:appenTOP}
\end{align}
The evaluation of the expectation values for the nearest-neighbours to the average vertex in the tree-triangle model is given by the following derivative 
\begin{equation}
     F_{\text{GCC}}' = \sum_{s'}\sum_{t'}\frac{d F_{\text{GCC}}}{d z_{s't'}}\bigg|_{z_{s't'}=1}
\end{equation}
where $F_{\text{GCC}}$ is given by unpacking Eq \ref{eq:F} for $\bm \tau=\{\bot,\Delta\}$ to find
\begin{align}
    {F}_{GCC}(\bm x,\bm y ) =&\ \sum_s\sum_t p_{s,t}f_\bot^{s}f_\Delta^{2t} -
    \sum_s\sum_t p_{s,t}g_\bot^{s}
   g_\Delta^{2t}
\end{align}
To evaluate this consider the following derivative 
 \begin{align}
     \frac{d F_{\text{GCC}}}{d z_{s't'}}\bigg|_{z_{s't'}=1} =&\  \frac{d}{d z_{s't'}}\bigg|_{z_{s't'}=1}G_0(f_{\bot},f_\Delta)-\frac{d}{d z_{s't'}}\bigg|_{z_{s't'}=1}G_0(g_{\bot},g_\Delta)\\
     =&\ \frac{d}{d z_{s't'}}\bigg|_{z_{s't'}=1}\sum_s\sum_tp_{st}f_{\bot}^sf_\Delta^{2t} -  \frac{d}{d z_{s't'}}\bigg|_{z_{s't'}=1}\sum_s\sum_tp_{st}g_{\bot}^sg_\Delta^{2t}\\
     =&\ \sum_s\sum_t p_{st}\left(
     sf_\bot^{s-1}\frac{df_\bot}{dz_{s't'}}f_\Delta^{2t} + 2tf_\bot^sf_\Delta^{2(t-1)}f_{\Delta}\frac{df_\Delta}{dz_{s't'}}
     \right) \nonumber\\
     &\ - \sum_s\sum_t p_{st}\left(
     sg_\bot^{s-1}\frac{dg_\bot}{dz_{s't'}}g_\Delta^{2t} + 2tg_\bot^sg_\Delta^{2(t-1)}g_\Delta\frac{dg_\Delta}{dz_{s't'}}
     \right)
 \end{align}
When evaluated at $z_{(s',t')}=1$ we have that $f_\tau(1)=1$ and so the first bracket simplifies significantly. The second bracket is more involved; however, using the self-consistent expressions for $u_\bot=G_{1,\bot}(u_\bot,u_\Delta^2)$ and $u_\Delta=G_{1,\Delta}(u_\bot,u_\Delta^2)$ we can write $g_{\bot}(1)=u_\bot$ and $g_\Delta(1)=u_\Delta$ to obtain
 \begin{align}
     \frac{d F_{\text{GCC}}}{d z_{s't'}}\bigg|_{z_{s't'}=1} =&\sum_s\sum_t p_{st}\left(
     sq_{\bot,(s',t')}
     + 2tq_{\Delta,(s',t')}
     \right)- \sum_s\sum_t p_{st}\big(
     su_\bot ^{s-1}q_{\bot,(s',t')}u_\bot^{s'-1}u_\Delta^{2t'}u_\Delta^{2t}\nonumber\\
     &+ 2tu_\bot ^su_\Delta^{2(t-1)}u_\Delta q_{\Delta,(s',t')}u_\bot^{s'}u_\Delta^{2(t'-1)}
     \big)
 \end{align}
 We now sum over $(s',t')$ to obtain
 \begin{align}
     \sum_{s'}\sum_{t'}\frac{d F_{\text{GCC}}}{d z_{s't'}}\bigg|_{z_{s't'}=1} =&\sum_s\sum_t p_{st}\left(
     s\sum_{s'}\sum_{t'}q_{\bot,(s',t')}
     + 2t\sum_{s'}\sum_{t'}q_{\Delta,(s',t')}
     \right)\nonumber\\
     &- \sum_s\sum_t p_{st}\bigg(
     su_\bot ^{s-1}u_\Delta^{2t}\sum_{s'}\sum_{t'}q_{\bot,(s',t')}u_\bot^{s'-1}u_\Delta^{2t'}\nonumber\\
     &+ 2tu_\bot ^su_\Delta^{2(t-1)}u_\Delta\sum_{s'}\sum_{t'}q_{\Delta,(s',t')}u_\bot^{s'}u_\Delta^{2(t'-1)}
     \bigg)
 \end{align}
 The probability distributions are normalised and hence have the following property $\sum_{s}\sum_t q_{\tau,(s,t)}=1$, so the first bracket reduces trivially to the sum of the average degrees of each edge topology. The second bracket also reduces; dealing first with the double summation over dashed variables we find 
  \begin{align}
     \sum_{s'}\sum_{t'}\frac{d F_{\text{GCC}}}{d z_{s't'}}\bigg|_{z_{s't'}=1} =&\sum_s\sum_t p_{st}\left(
     s
     + 2t\right) - \sum_s\sum_t p_{st}\left(
     su_\bot^{s-1}u_\Delta^{2t}
     u_\bot + 2tu_\bot ^su_\Delta^{2(t-1)}u_{\Delta}^2
     \right)
 \end{align}
 before observing that 
 \begin{align}
     \sum_s\sum_tp_{st}sx^{s-1}y^{t} = \langle s\rangle G_{1,\bot}(x,y)\\
     \sum_s\sum_tp_{st}tx^{s}y^{t-1} = \langle t\rangle G_{1,\Delta}(x,y)
 \end{align}
 to arrive at 
 \begin{equation}
      \sum_{s'}\sum_{t'}\frac{d F_{\text{GCC}}}{d z_{s't'}}\bigg|_{z_{s't'}=1} =
      \langle s\rangle +2\langle t\rangle - \langle s\rangle G_{1,\bot} (u_\bot , u_\Delta^{2})u_\bot - 2\langle t\rangle G_{1,\Delta} (u_\bot , u_\Delta^{2})u_\Delta^2
 \end{equation}
Substituting the self-consistent relationships for $u_\bot$ and $u_\Delta$ we finalise the expression as
\begin{equation}
      \sum_{s'}\sum_{t'}\frac{d F_{\text{GCC}}}{d z_{s't'}}\bigg|_{z_{s't'}=1} =
      \langle s\rangle \left(1 - u_\bot^2\right)
      +
      2\langle t\rangle\left(1 -  u_\Delta^3\right)\label{eqAppenBOT} 
 \end{equation}
 In the case that there are no triangles present in the model, then $u_\Delta=1$ and $\langle t\rangle =0$; the expression reduces to 
 \begin{align}
     \sum_{s'}\frac{d F_{\text{GCC}}}{d z_{s'}}\bigg|_{z_{s'}=1} =\langle s\rangle \left(1 - u_\bot^2\right)
 \end{align}
 which is the result of \cite{Mizutaka_2020} in the case that $l=1$. In the opposite case, when there are no ordinary edges, we find
 \begin{equation}
     \sum_{t'}\frac{d F_{\text{GCC}}}{d z_{t'}}\bigg|_{z_{t'}=1} =2\langle t\rangle \left(1 - u_\Delta^3\right)
 \end{equation}
The probability $P(k_{\tau,0},k_{\tau,1}) = P((s_0,t_0), (s',t'))$ is given by the quotient of Eqs \ref{eq:appenTOP} and \ref{eqAppenBOT} where we find
\begin{align}
   P((s_0,t_0), (s',t')) =&\ \frac{d \hat{F}_{\text{GCC}}}{d z_{s't'}}\bigg|_{z_{s't'}=1}\bigg/\sum_{s'}\sum_{t'}\frac{d F_{\text{GCC}}}{d z_{s't'}}\bigg|_{z_{s't'}=1}\nonumber\\
      =&\ p_{s_0t_0}\bigg(s_0q_{\bot,(s',t')} + 2t_0q_{\Delta,(s',t')}\bigg)- p_{s_0t_0}\bigg(s_0u_\bot^{s_0-1}q_{\bot,(s',t')}u_\bot^{s'-1}u_\Delta^{2t'}u_\Delta^{2t_0}\nonumber\\
     &+ 2t_0u_\bot ^{s_0}u_\Delta^{2(t_0-1)}u_\Delta q_{\Delta,(s',t')}u_\bot^{s'}u_\Delta^{2(t'-1)}
     \bigg) \bigg/ \langle s\rangle \left(1 - u_\bot^2\right)
      +
      2\langle t\rangle\left(1 -  u_\Delta^3\right)
\end{align}
The conditional probability that a neighbour has joint degree $(s',t')$ given a focal vertex of joint degree $(s_0,t_0)$ is 
\begin{equation}
    P(s',t'\mid s_0,t_0) =\frac{p_{s_0t_0}s_0q_{\bot,(s',t')}\left(1-u_\bot^{s_0+s'-2}u_\Delta^{2(t_0+t')}\right)+ 2p_{s_0t_0}t_0q_{\Delta,(s',t')}\left(1-u_\bot^{s_0+s'}u_\Delta^{2(t_0+t'-2)+1}\right)}{p_{s_0t_0}(s_0+2t_0)\left(1-u_\bot^{s_0}u_\Delta^{2t_0}\right)}
\end{equation}
Using Eq \ref{eq:average_deg} we find the average joint degree of a neighbour to a $(s_0,t_0)$ vertex as
\begin{equation}
   \mathcal E[\bm k_{\bm \tau,1}\mid \bm k_{\bm \tau,0}]= \left(\sum_{s',t'}s' P(s',t'\mid s_0,t_0),\sum_{s',t'}t' P(s',t'\mid s_0,t_0)\right)^T\label{eq:avetreetriangle}
\end{equation}

\begin{figure}
\begin{center}
\includegraphics[width=0.40\textwidth]{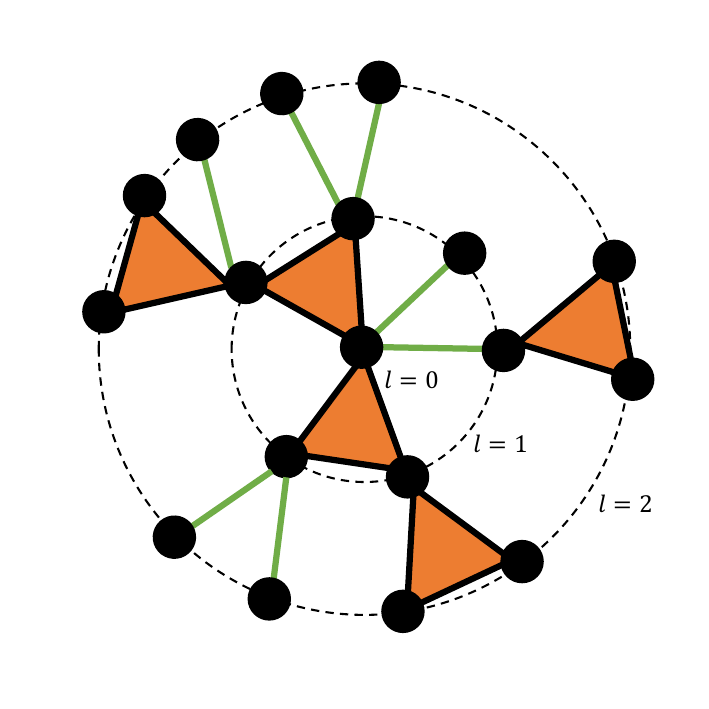}
\caption[correlations]{ 
An example of the degree correlation model in the tree-triangle model; 3-cliques are shaded orange whilst 2-cliques are coloured green. The joint degree of the focal vertex in layer $l=0$ is $k_{\bm \tau,0} = (2,2)$. We can follow edges of topology $\bot$ or $\Delta$ to the first neighbours. The distribution of the joint degrees of vertices in layer $l=2$ depends on the topology of the path that we choose to reach it. Note, we do not traverse edges between triangles that lead to vertices in the same layer.
} \label{fig:correlations}
\end{center}
\end{figure}

\end{widetext}

\appendix

\end{document}